\documentclass[12pt]{article}

\usepackage[margin=2.5cm]{geometry}
\usepackage[usenames,dvipsnames]{xcolor}
\definecolor{SubtleColor}{rgb}{0,0,.50}
\usepackage[letterpaper=true,colorlinks=true,pdfpagemode=none,allcolors=SubtleColor,pdfstartview=FitH,backref=page]{hyperref}
\usepackage{graphicx}
\usepackage[authoryear]{natbib}

\usepackage{amsmath,amsfonts,amssymb}
\usepackage{mathtools}
\usepackage{algorithm,algorithmic}

\usepackage{pgf}
\usepackage{tikz}
\usetikzlibrary{fit,arrows,automata,shapes}					% fitting shapes to coordinates
\usetikzlibrary{backgrounds}
\tikzstyle{state}=[circle,thick,minimum size=1.2cm, draw=black]
\tikzstyle{measurement}=[circle,thick,minimum size=1.2cm,draw=black,
fill=gray!25]  
\tikzstyle{switch}=[rectangle,thick, minimum size=1cm, draw=black]

\newcommand{\blkdiamond}{\raisebox{0pt}{\tikz{\node[draw,scale=0.4,diamond,fill=black](){};}}}

\usepackage{wasysym}

\usepackage{booktabs}

\graphicspath{{gfx/}}%{../figs-for-paper_files/figure-latex/}}
\usepackage{mathtools}
\mathtoolsset{showonlyrefs,showmanualtags}

\renewcommand{\hat}{\widehat}
\usepackage{upgreek}
\DeclareRobustCommand{\varx}{{\mathpalette\irchi\relax}}
\newcommand{\irchi}[2]{\protect\raisebox{\depth}{$#1\upchi$}}
\newcommand{\given}{\ \vert\ }
\newcommand{\E}{E}
\newcommand{\Expect}[1]{\E\left[#1\right]}
\newcommand{\Var}[1]{{Var}\left[#1\right]}

\newcommand{\email}[1]{\href{mailto:#1}{#1}}
\renewcommand{\top}{\mathsf{T}}
\usepackage{authblk}

\begin{document}

\title{Markov-Switching State Space Models for Uncovering Musical
  Interpretation}

  \author[a,1]{Daniel J.\ McDonald}
  \author[b]{Michael McBride}
  \author[c]{Yupeng Gu}
  \author[c]{Christopher Raphael}
  
  \affil[a]{Department of Statistics, The University of British Columbia}
  \affil[b]{Department of Statistics, Indiana University}
  \affil[c]{School of Informatics, Computing, and Engineering, Indiana University}
  
\date{Last updated: 1 September 2021}
\maketitle

\footnotetext[1]{To whom correspondence should be
  addressed. E-mail: \email{daniel@stat.ubc.ca}. This work began while DJM was a
  member of the Department of Statistics at Indiana University. This work was partially supported by 
  the National Science Foundation Grants DMS-1407439
  and DMS-1753171 (to DJM) and Grant IIS-1526473 (to CR).} 

\begin{abstract}
  For concertgoers, musical interpretation is the most important factor
  in determining whether or not we enjoy a classical performance. Every
  performance includes mistakes---intonation issues, a lost note, an
  unpleasant sound---but these are all easily forgotten (or unnoticed) when a performer
  engages her audience, imbuing a piece with novel emotional content
  beyond the vague instructions inscribed on the printed page.
  % While music teachers use
  % imagery or heuristic guidelines to motivate interpretive decisions, combining these
  % vague instructions to create a convincing performance remains the domain
  % of the performer, subject to the whims of the moment, technical
  % fluency, and taste.
  In this
  research, we use data from the CHARM Mazurka Project---forty-six professional
  recordings of Chopin's Mazurka Op.\ 68 No.\ 3 by consummate
  artists---with the goal of 
  elucidating musically interpretable performance decisions. We
  focus specifically on each performer's use musical tempo by
  examining the
  inter-onset intervals of the note attacks in the recording. To
  explain these tempo decisions, we
  develop a switching state space model and estimate it by maximum
  likelihood combined with
  prior information gained from music theory and performance
  practice. We use the estimated
  parameters to quantitatively describe individual performance
  decisions and compare recordings. These comparisons suggest methods
  for informing music 
  instruction, discovering listening preferences, and analyzing performances.
  
  % The text of your abstract. 200 or fewer words. At 172 currently
\end{abstract}

\tableofcontents

\section{Introduction}
\label{sec:introduction}

% \attn{See \href {http://amstat.tfjournals.com/asa-style-guide/}{here}
%   for the style guide (detailed) and
%   \href{https://www.tandfonline.com/action/authorSubmission?journalCode=uasa20&page=instructions}{here}
%   for the author instructions.}

Statistical analysis of the musical content of recordings has
become more and more important to academics and industry. Online
music services like Pandora, Last.fm, Spotify, and others rely on
recommendation systems to suggest potentially interesting or related
songs to listeners. In 2011, the KDD Cup challenged academic computer
scientists and statisticians to identify user tastes in music with the
\href{http://labrosa.ee.columbia.edu/millionsong/}{Yahoo! Million
  Song Dataset} (see \citet{DrorKoenigstein2012} for details of the
competition). Pandora, through its proprietary
\href{https://www.pandora.com/about/mgp}{Music Genome Project}, uses
trained musicologists to assign new songs a vector of trait
expressions (consisting of up to 500 ``genes'' depending on the genre)
which can then be used to measure similarity with other
songs. However, most of this work has focused on the analysis of more popular
and more profitable genres of music---pop, rock, country---as opposed
to classical music. 

Western classical music is a subcategory
whose boundaries are occasionally difficult to define. But
the distinction is of great importance when it comes to the analysis
which we undertake here. Leonard Bernstein, the great composer,
conductor, and pianist, gave the following characterization in one of his famous
``Young People's Concerts''
broadcast by the Columbia Broadcasting Corporation in the 1950s and
1960s \citep{Bernstein2005}.
% \begin{quote}
%   You see, everybody thinks he knows what classical music is: just any music that isn't jazz,
%   like a Stan Kenton arrangement or a popular song, like ``I Can't Give
%   You Anything but Love Baby,'' or folk music, like an African war
%   dance, or ``Twinkle, Twinkle Little Star.'' But that isn't what
%   classical music means at all.
% \end{quote}
% Bernstein goes on to discuss an important distinction between what
% we often call ``classical music'' and other types of music which is
% highly relevant to the current study.
\begin{quote}
  %The real difference is that
  [\ldots W]hen a composer
  writes a piece of what's usually called classical music, he puts down
  the exact notes that he wants, the exact instruments or voices that he
  wants to play or sing those notes [\ldots]
  %---even the exact number of
  %instruments or voices;
  and he also writes down as many directions as
  he can think of. [\ldots] Of course, no performance can be perfectly exact, because there
  aren't enough words in the world to tell the performers everything
  they have to know about what the composer wanted.
  % But that's just what
  % makes the performer's job so exciting---to try and find out from what
  % the composer did write down as exactly as possible what he meant. Now
  % of course, performers are all only human, and so they always figure it
  % out a little differently from one another.  
\end{quote}
What separates classical music from other types of music is that the
music itself is written down very precisely but performed millions of times in a
variety of interpretations.\footnote{Of course jazz, rock, pop and
  other genres are often written down for future performances, but
  without the same precision. Compare for example a lead sheet John
  Coltrane's ``Giant Steps'' to his original 1960 recording.}
There is no ``gold standard'' recording to
which everyone can refer but rather a document created for
reference. Therefore, the musical genome technique mentioned above
will only relate ``pieces'' but not ``performances''. We need
new methods to decide whether we prefer Leonard Bernstein's
recording of Beethoven's Fifth Symphony or Herbert von Karajan's and
to articulate why.

In this paper, we develop a statistical model for some of the
decisions that a musician must make for classical music
interpretations. We focus on how the musician modulates
{\it tempo}, or speed, over the course of a recording. 
\begin{figure}[t]
  \centering
  \includegraphics[width=.9\linewidth]{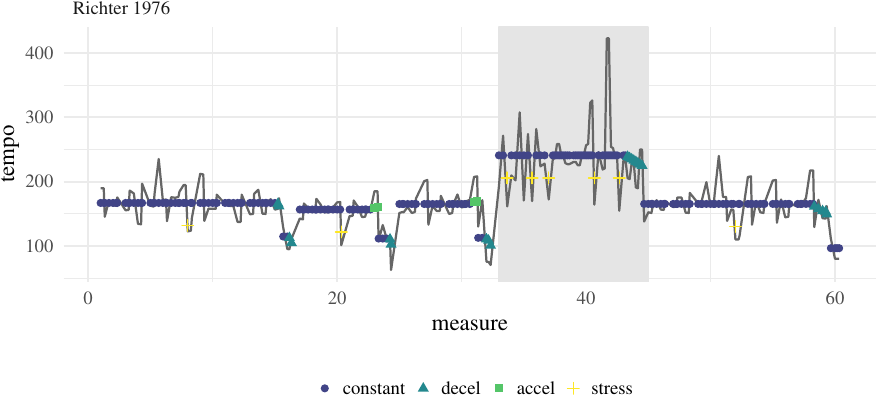}
  \caption{Note-by-note tempos for a recording of Chopin's Mazurka
    Op.\ 68 No.\ 3 by Sviatoslav Richter. The solid line are the
    observed tempos, while the dots represent inferred tempo states
    from our model. }
  \label{fig:richter}
\end{figure}
\autoref{fig:richter} shows the tempo\footnote{Technically, by
  ``tempo'', we mean the ratio of musical time to clock time as $0.25$
  beats $/$ $0.1$ seconds $=$ 150 beats per minute. A musician would
  likely think of ``tempo'' more broadly as something like a ``typical
  speed regime'' akin to the ``constant tempo'' state we use in our
  decision model below. We will not generally distinguish between
  these two interpretations and use the more succinct ``tempo''.}
  in beats-per-minute (b.p.m.) of
a recording made by Sviatoslav Richter of Chopin's Mazurka Op.\ 68
No.\ 3. The solid line shows the actual tempo at which he plays each
note, while the points correspond to our model's inferences
for his actual intentions. Some of this intent is prescribed  by
Chopin in his music, but the extent to which Richter observes Chopin's
indications makes his recording different from those of other
pianists. It is these differences that we hope to capture and understand.

\subsection{Related work}
\label{sec:related-work}

The vast majority of work at the intersection of statistics
%or machine learning
and classical music analysis has focused on a handful of tasks,
most notably structure analysis, music generation, and score
alignment.

Analysis of musical structure and its relationships with interpretation
forms the basis of music theory, and hence constitutes the core of standard
conservatory curricula, along with history and performance. Automatically discovering musical structures
from performances without expert input has become more relevant
recently. \citet{RenDunson2010} use Dirichlet process models to identify
similar sections of individual classical music
performances. \citet{RobertsEngel2018} use variational autoencoders to
discover long-term structure with an explicit goal toward improved
automatic music composition.

Computer music generation and composition has a long
history~\citep{SturmBen-Tal2019,Boulanger-LewandowskiBengio2012,Collins2016,Ariza2005,FlossmannGrachten2013}.
It is actively investigated, especially using deep
learning~\citep{HadjeresPachet2017}, and has become commercially
relevant for advertising and video games through companies like
Aiva (\url{aiva.ai}), and Melodrive (\url{melodrive.com}). Google has
developed the Magenta project to enable open-source music
composition~\citep{RobertsHawthorne2018}.

The score alignment problem matches live or recorded performances to
the musical score, a necessary processing step for any 
automated analysis. On-line alignment processes audio waveforms in
real-time and is sometimes called score
following~\citep{DannenbergRaphael2006,Cont2010,ContSchwarz2007,ArztWidmer2015}. Audio
matched to the score can then be used as an input for automated
musical accompaniment~\citep{Raphael2010,Vercoe1985,Dannenberg1985}.
Given recorded accompaniment, these systems modulate playback
in response to a live soloist who both makes interpretive timing
decisions and mistakes. Off-line alignment~\citep{Earis2007} can be
used to analyze the recordings, as we do here, or for
generating descriptive features of the
performance~\citep{ThickstunHarchaoui2017}, possibly for later
analysis in recommender
systems~\citep{McFeeLanckriet2011,OordDieleman2013}.
For an overview of %these and
related goals in music information retrieval,
see~\citet{schedl2014music}.

More closely connected to the work here is the literature
on expressive synthesis \citep{Grindlay, Flossman, Widmer, Maezawa,
Bresin, Arcos}.  In this domain, one seeks to create a musically
satisfying performance of a score, often given a training set of
score-performance pairs.
% that can be used as a learning set or as the
% basis for case-based reasoning.
This problem is highly challenging
since musical interpretation relies on latent aspects of the music,
such as structure, stress, grouping, closure, surprise, affect,
and others not explicitly appearing in the score.  Music theorists often
describe the score as the {\em surface} of the music, thus recognizing
that there is much that lies beneath this surface.  Though we do not
treat expressive synthesis, our work shares the need for explicitly
representing expressive performances.% found in this domain.

Most recent work in expressive synthesis is based in machine
learning, both in terms of methodology and the agnostic spirit of the
modeling.  Here one parametrizes the performance in terms of variables
that will be estimated from the score, such as rate of change of tempo
or dynamics.  For instance, \citet{Flossman} represent the joint
configuration of local performance parameters and score parameters as
a conditional Gaussian distribution, learned from training, and then
used to estimate the expression on a new score.  We also use
conditional Gaussians as a modeling element, though we try to model
the {\em process} nature of the music, rather than viewing each note
independently.  The work on expressive synthesis of
\citet{Maezawa} uses an autoregression to model the expressive
parameters of a note, estimating the autoregressive parameters using
deep learning on score attributes.  The use of autoregression
accomodates the smoothness normally found in expressive performance,
while still not being overly prescriptive about the nature of the
musical evolution.

Common to these approaches are the generic assumptions relating the
musical score and the expressive performance parameters, hoping to
push the hardest work --- understanding what is important --- onto the
learning algorithm.  In this spirit one seeks to discover what
relevant sub-surface attributes of the score can be correlated with
performance decisions.  Perhaps the most extreme example of this
musically agnostic approach is \citet{Grindlay}, who use a
minimally-primed hidden Markov model to generate expression without
giving any explicit meaning to the states.

We emphasize an important
modeling difference between these approaches and what we propose.  We
explicitly model the performance in terms of a switching Kalman
filter, thus making rather strong {\em a priori} assumptions derived
from our musical sensibilities.
% In this regard these is a close
% connection between our work and that of
\citet{Chew} similarly use an
explicit parametrization of tempo evolution by automatically
partitioning the music into segments represented by local quadratics.
Our appoach considers a broader family of parametrizations but shares
the basic approach of {\em building in} musical knowledge to the
model.

\subsection{Our contributions}
\label{sec:our-contributions}

In this paper, we develop a switching Kalman filter model for the
tempo decisions a performer makes in recorded classical music. We
present an algorithm for performing likelihood inference, estimate our
model using a large collection of recordings of the same composition,
and demonstrate how the model is able to recover performer intentions,
and how they relate to standard musical analysis. We use the 
low-dimensional representations to compare and contrast the recordings, and 
discuss how this analysis facilitates more informed musical
comparisons of the recordings. Such an analysis may help listeners to
choose other performers whose tendencies are similar (or dramatically
different!) from those they already enjoy, suggest new recordings to
purchase, or motivate future concert attendance behaviors. Our
analysis can also aid automatic performance generation by reusing a
musician's estimated parameters on a reproducing instrument or
potentially inform
music education. In combination with other software, a music teacher
could create visuals for a student's performance, such as those
presented in this paper, and directly discuss areas for improvement.

In \autoref{sec:materials-methods} we discuss our dataset, a
collection of professional recordings of Chopin's Mazurka Op.\ 68 No.\
3. We also present our model for tempo decisions, discuss its
statistical estimation, and detail its utility for understanding
interpretations as a musician would. \autoref{sec:analys-chop-mazurka}
presents a music theory interpretation of the Mazurka. We discuss how
different performers approach this piece through the lens of our model.
We also examine groups of performances based on our model and
interpret the musical 
meaning of these groupings.  Finally, we contrast our approach with
some alternative non-parametric
smoothers, discusses their deficiencies relative our switching model,
and examine some issues with our proposal.

\section{Materials and methods}
\label{sec:materials-methods}

%\subsection{Data and preprocessing}

In this paper, we examine note-by-note tempos for 46 recordings of
Chopin's Mazurka Op.\ 68, No.\ 3. The data is part of a large
collection of the complete Chopin Mazurkas and other recordings
assembled and analyzed by the Center for the History and Analysis of
Recorded Music (CHARM) in the United Kingdom~\citep{CHARM-site}. The
recordings were processed using the note-onset detection algorithm
developed by~\citet{Earis2007} and are available for
download~\citep{Earis2009}. We use the data for ``all rhythmic
events'', which includes the time of each note attack as well as it's
relative loudness.
%For the sake of reproducibility, we have included this data in our R package.

\subsection{Switching state-space models}
\label{sec:switch-state-space}

State-space models define the probability distribution of a continuous
time series $Y$ by reference to some imagined, continuous hidden state, $X$. In
particular, the observation at time $i$ is assumed to be
independent of past and future observations conditional on the state
at time $i$. Coupling with temporal dependence for $X$---most
frequently obeying the Markov property---induces a temporal model for
the observations.  The most general form of a state-space model is
then characterized by the 
measurement equation (the conditional probability of observations
given the states),
the transition equation (specifying the nature of Markovian
dynamics), and an initial distribution for the state: 
\begin{equation}
\begin{aligned}
  y_i &= f_\theta(x_i,\epsilon_i), &
  x_{i+1} &= g_\theta(x_i,\eta_i), &
  x_1 &\sim F,
\end{aligned}
\label{eq:ssmod}
\end{equation}
where $\epsilon_i$ are $\eta_i$ are marginally and mutually independent and $F$ is
an arbitrary but specified distribution. Both
$y_i$ and $x_i$ can be vector-valued, though in our
application, $y_i$ will be univariate. The
vector $\{y_i\}_{i=1}^n$ is observed, and the goal is to make
inferences for the unobserved states $\{x_i\}_{i=1}^n$ as well as any
unknown parameters $\theta$ characterizing $f_\theta$, $g_\theta$, and
the distributions of $\epsilon_i$ and $\eta_i$.
% \autoref{fig:ssmod}
% shows a directed acyclic graph for the dependence structure in the
% typical state-space model.

% \begin{figure}
%   \centering
%   % The continuous state vector is represented by a circle.
%   % "minimum size" makes sure all circles have the same size
%   % independently of their contents.
%   \begin{tikzpicture}[>=latex,text height=1.5ex,text
%     depth=0.25ex,ampersand replacement=\&] 
%     \matrix[row sep=1cm,column sep=1cm] {
%       % first line: hidden continuous state
%       \node (x_k-2) {$\cdots$}; \&
%       \node (x_k-1) [state] {$\mathbf{x}_{i-1}$}; \&
%       \node (x_k)   [state] {$\mathbf{x}_i$};     \&
%       \node (x_k+1) [state] {$\mathbf{x}_{i+1}$}; \&
%       \node (x_k+2) {$\cdots$};
%       \\
%       % Second line: Measurement
%       \node (y_k-2) {$\cdots$}; \&        
%       \node (y_k-1) [measurement] {$\mathbf{y}_{i-1}$}; \&
%       \node (y_k)   [measurement] {$\mathbf{y}_i$};     \&
%       \node (y_k+1) [measurement] {$\mathbf{y}_{i+1}$}; \&
%       \node (y_k+2) {$\cdots$};
%       \\
%     };
%     % The diagram elements are now connected through arrows:
%     \path[->]
%     (x_k-2) edge (x_k-1)
%     (x_k-1) edge (x_k)	
%     (x_k)   edge (x_k+1)	
%     (x_k+1)   edge (x_k+2)	
%     (x_k-1) edge (y_k-1)
%     (x_k) edge (y_k)
%     (x_k+1)   edge (y_k+1);
%   \end{tikzpicture}
%   \caption{State-space model. Filled objects are observed, circles
%     indicate that both hidden and observed states are
%     continuous.\label{fig:ssmod}} 
% \end{figure}

If $f_\theta$ and $g_\theta$ are linear, and $\epsilon_i$,
$\eta_i$, and $F$ assumed to have Gaussian distributions, Equation \eqref{eq:ssmod} becomes
\begin{equation}
  \begin{aligned}
    x_{i+1}&= d+T x_i + \eta_{i}, 
    & \eta_i &\sim N(0,\ Q),     
    &x_1 &\sim N(x_0,\ P_0),\\
    y_i&= c + Z x_i + \epsilon_i,     
    & \epsilon_i &\sim N(0,\ G), \\
  \end{aligned}
  \label{eq:lgmod}
\end{equation}
where the vectors $c,\ d$ and matrices $T,\ Z,\ Q,$ and $G$ are allowed to depend
on $\theta$, can potentially vary with $i$, or can depend on previous
values of $x$ and $y$. In this case,
the Kalman filter %\autoref{alg:kalman}
\citep[see for example,][]{Kalman1960,Harvey1990},
provides closed form 
solutions for the conditional
distributions of the states and gives the likelihood of $\theta$
given data. For completeness, we have included the Kalman filter
inference algorithm in
the Supplement \citep{McDonaldMcBride2021a}.% as \autoref{alg:kalman}.

Although the Kalman filter %\autoref{alg:kalman}
returns the likelihood for $\theta$, and is therefore all we need for
parameter estimation,
inference for the mean and variance
of $X$ is conditional only on the preceding observations
$\{y_j\}_{j=1}^i$: $\varx_i=\Expect{x_i\given y_1\ldots,y_i}$ and
$P_i=\Var{x_i\given y_1,\ldots,y_i}$. To
incorporate all future observations into these estimates, and produce
the inferred performance tempos shown in, for example,
\autoref{fig:richter}, the Kalman smoother is required. 
% \begin{algorithm}[t!]
%   \begin{singlespace}
%   \caption{Kalman smoother (Rauch-Tung-Striebel): estimate $\hat{X}$ conditional on
%     $Y$\label{alg:kalman-smoother}} 
%   \begin{algorithmic}
%     \STATE {\bf Input:} $\varx$, $\widetilde{X}$, $P$, $\widetilde{P}$,
%     $T,$ $c$, $Z$.
%     \STATE $t=n$,
%     \STATE $\hat{x}_{n}\leftarrow \widetilde{x}_n$, 
%     \WHILE{$t>1$}
%     \STATE $\hat{y}_i \leftarrow c + Z\hat{x}_i,$
%     \COMMENT{Predict observation vector}
%     \STATE $\begin{aligned} e &\leftarrow \hat{x}_i -
%       \varx_i, & V &\leftarrow P_i^{-1}\end{aligned}$,
%     \STATE $t\leftarrow i-1$, \COMMENT{Increment}
%     \STATE $\hat{x}_i = \widetilde{x}_i + \widetilde{P}_i T Ve $ 
%     \ENDWHILE
%     \RETURN $\widehat{Y}=\{\hat{y}_i\}_{i=1}^n, \hat{X}=\{\hat{x}_i\}_{i=1}^n$
%   \end{algorithmic}
% \end{singlespace}
% \end{algorithm}
Many different smoother algorithms have been tailored for different
applications. %\autoref{alg:kalman-smoother} (Supplementary Material),
The smoother we use, due
to~\citet{RauchStriebel1965}, is often referred to as the classical
fixed-interval smoother~\citep{AndersonMoore1979}. It produces only
the unconditional expectations of the hidden state
$\hat{x}_i=\Expect{x_i\given y_1,\ldots,y_n}$, which is all that is
necessary for our analysis. This algorithm is again given in
the Supplementary Material \citep{McDonaldMcBride2021a}. We note that the smoother estimate of
$\Var{x_i\given y_1,\ldots,y_n}$ is not necessary for any of the
analysis discussed in this paper.

Linear Gaussian state-space models can be made quite flexible
by expanding the state vector or allowing the parameter matrices to
vary with time. Furthermore, this general form encompasses many
standard time series models: ARIMA models, ARCH and GARCH models,
stochastic volatility models, exponential smoothers, and
more~\citep[see][for many other
examples]{DurbinKoopman2001}. Nonlinear, non-Gaussian versions have
been extensively
studied~\citep{DurbinKoopman1997,Fuh2006,Kitagawa1987,Kitagawa1996}
and algorithms for filtering, smoothing, and parameter estimation have
been derived~\citep[for example,][]{KoyamaPerez-Bolde2010,AndrieuDoucet2010}. 
However, these models are less useful
for change-point detection or other discontinuous behavior
when the times of discontinuity are unknown. 

To remedy this deficiency, one can use a switching state-space
model as shown in \autoref{fig:switchss}. Here, we assume $\{s_i\}_{i=1}^n$ is a
hidden, discrete process with Markovian dynamics. Then, the value of
the hidden state at time $i$, $s_i=k$ say, can determine the evolution of
the continuous model at time $i$. The graphical model in
\autoref{fig:switchss} gives the conditional independence properties
we will use in our model for musical interpretation, 
representing just one of many possiblities. Switching state-space models have a long
history with applications ranging from
economics~\citep{KimNelson1998,Kim1994,Hamilton2011} to speech
processing~\citep{FoxSudderth2011} to animal
movement~\citep{PattersonThomas2008,BlockJonsen2011}. \citet{GhahramaniHinton2000}
provide an excellent overview of the history, 
typography, and algorithmic developments. In Equation~\eqref{eq:lgmod}, the
parameter matrices were not time varying. We
allow the switch states $s_i, s_{i-1}$, along with the parameter
vector $\theta$, to determine the specific dynamics at time $i$:
\begin{equation}
  \begin{aligned}
    x_1 &\sim N(x_0,\ P_0),\\
    x_{i+1}&= d(s_i,s_{i-1})+T(s_i,s_{i-1}) x_i + \eta_i, 
    & \eta_i &\sim N(0,Q(s_i,s_{i-1})),\\
    y_i&= c(s_i) + Z(s_i) x_i + \epsilon_i, & \epsilon_i &\sim N(0, G(s_i)).
  \end{aligned}
\end{equation}
In other words, the hidden Markov (switch) state determines the
collection of $d,\ c,\ T,\ Z,$ $G$ and $Q$
that govern the evolution of the system. Allowing $d$, $T$, and $Q$,
to depend on $s_{i-1}$ in addition to $s_i$ (the diagonal arrow in
\autoref{fig:switchss}) will allow us to incorporate acceleration as
well as velocity into our model for tempo decisions.

\begin{figure}
  \centering
  \begin{tikzpicture}[>=latex,text height=1ex,text depth=0.25ex,ampersand replacement=\&]
    % The various elements are conveniently placed using a matrix:
    \matrix[row sep=.5cm,column sep=.5cm] {
      % First line: Switch state
      \node (s_k-2)  {$\cdots$}; \&
      \node (s_k-1) [switch]{$s_{i-1}$}; \&
      \node (s_k)   [switch]{$s_i$};     \&
      \node (s_k+1) [switch]{$s_{i+1}$}; \&
      \node (s_k+2) {$\cdots$};
      \\
      % Second line: hidden continuous state
      \node (x_k-2) {$\cdots$}; \&
      \node (x_k-1) [state] {$\mathbf{x}_{i-1}$}; \&
      \node (x_k)   [state] {$\mathbf{x}_i$};     \&
      \node (x_k+1) [state] {$\mathbf{x}_{i+1}$}; \&
      \node (x_k+2) {$\cdots$};
      \\
      % Third line: Measurement
      \node (y_k-2) {$\cdots$}; \&        
      \node (y_k-1) [measurement] {$\mathbf{y}_{i-1}$}; \&
      \node (y_k)   [measurement] {$\mathbf{y}_i$};     \&
      \node (y_k+1) [measurement] {$\mathbf{y}_{i+1}$}; \&
      \node (y_k+2) {$\cdots$};
      \\
    };
    % The diagram elements are now connected through arrows:
    \path[->]
    (s_k-2) edge (s_k-1)
    (s_k-1) edge (s_k)	
    (s_k)   edge (s_k+1)	
    (s_k+1)   edge (s_k+2)	
    
    (x_k-2) edge (x_k-1)
    (x_k-1) edge (x_k)	
    (x_k)   edge (x_k+1)	
    (x_k+1)   edge (x_k+2)	
    
    (s_k-1) edge (x_k-1)
    (s_k) edge (x_k)
    (s_k+1)   edge (x_k+1)
    
    (x_k-1) edge (y_k-1)
    (x_k) edge (y_k)
    (x_k+1)   edge (y_k+1)
    
    (s_k-1) edge (x_k)
    (s_k) edge (x_k+1)
    (s_k+1)   edge (x_k+2)
    
    (s_k-1) edge[bend left] (y_k-1)
    (s_k) edge[bend left] (y_k)
    (s_k+1)   edge[bend left] (y_k+1)
    ;
  \end{tikzpicture}
  \caption{Switching state space model. Filled objects are observed,
    rectangles are discrete, and circles are continuous.\label{fig:switchss}}
\end{figure}

\subsection{A model for tempo decisions}

In musical scores, {\em tempi} (the Italian plural of tempo) may be
marked at various points throughout a piece of music. The
beginning can be either explicit, with a metronome marking to
indicate the number of beats per minute (b.p.m.), and/or with some words
(e.g., {\em Adagio}, {\em Presto}, {\em Langsam}, Sprightly) which indicate an
approximate speed. 
\begin{figure}[t!]
  \centering
  \includegraphics[height=2cm]{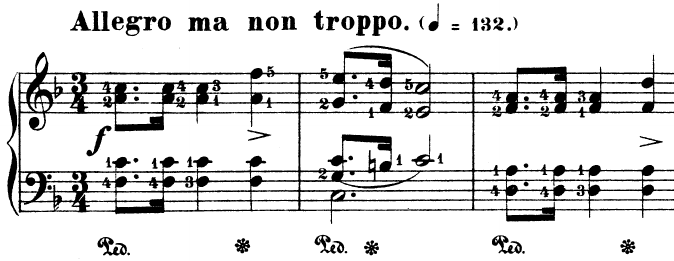}
  \includegraphics[height=2cm]{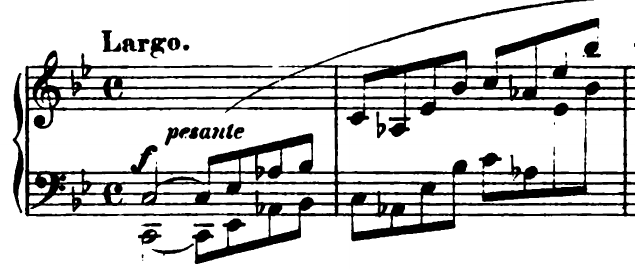}
  \caption{The beginning of two Chopin piano compositions: the Mazurka
    we analyze is on the left while the Ballade No.\ 1, Op.\ 23 is on
    the right.}
  \label{fig:tempo-markings}
\end{figure}
\autoref{fig:tempo-markings} shows the beginning of two Chopin piano
compositions: the Mazurka we analyze and the Ballade No.\ 1, Op.\
23. The initial tempo of the Mazurka is given with a metronome
marking as well as the Italian phrase {\em Allegro ma non troppo}
(``cheerful, but not too much''). The beginning of the Ballade is 
marked {\em Largo}, which translates literally as ``broad'' or
``wide'', and modified by the stylistic indication {\em pesante}
(``heavy''). Obviously, the metronome markings are much more exact,
though even these are often viewed as suggestions rather than
commandments. The metronome markings in most of Beethoven's
compositions, for example, are notoriously fast, and some scholars
believe that his metronome (one of the first ever made) was
inaccurate~\citep{ForsenGray2013}. Often, compositions will have numerous such markings later
in the piece of music, but these are only some of the ways that tempo
is indicated. Composers will also indicate periods of speeding-up
(\emph{accelerando}) or
slowing-down (\emph{ritardando}).

\begin{figure}[t!]
 \centering
 \includegraphics[width=.9\linewidth]{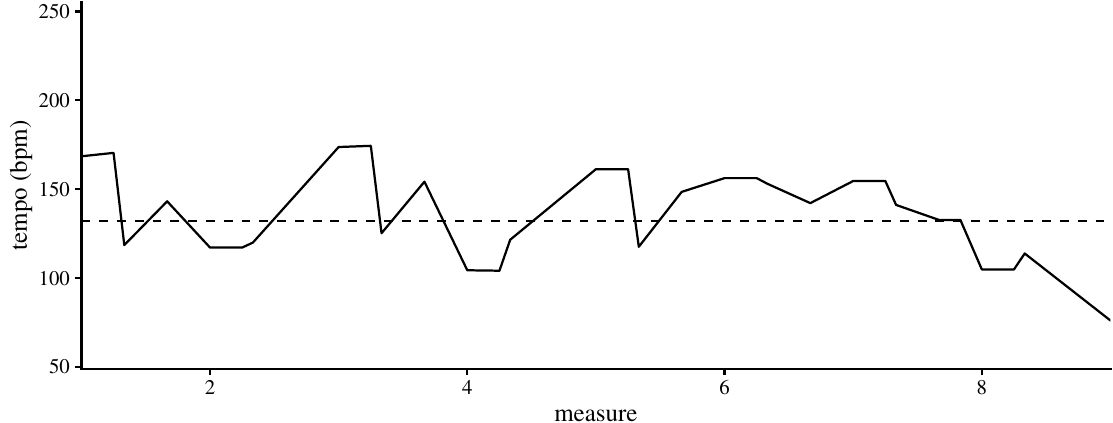}
 \caption{The solid line shows the observed note-by-note tempo for
   the beginning of the Mazurka as performed by Arthur Rubinstein in
   1961. The dashed line indicates 132 b.p.m.}
 \label{fig:short-perf}
\end{figure}

Absent instructions from the composer, performers generally maintain
(or try to maintain) a steady tempo, and this assumption plays a major
role in our model of tempo decisions. Of course, a normal human 
never plays precisely like a 
metronome, although she may try quite hard to do so. The observed
ratio of musical time to clock time
can therefore be thought of as stochastic, the sum of an
intentional, constant tempo, plus noise representing inaccuracy
or, perhaps more charitably, unintentional variation which the
listener fails to perceive as ``wrong''.\footnote{Some may argue with
  this explanation. As one anonymous reviewer pointed out, describing
  deviations from constant tempo as ``unintentional noise'' fails to
  account for the possibility that the performer is consciously or
  unconsciously controlling these small deviations, and their success
  as artists can be partially attributed to their preternatural
  abilities to exert such control. In the end, our
  model is smoothing such deviations away for the sake of providing a
  low-dimensional explanation of performance behavior. See
  \autoref{sec:this-model-reas} for a discussion of how much might be
  lost by this smoothing.} For instance, the example in
\autoref{fig:short-perf} shows the beginning of the piece as performed
by Arthur Rubinstein in a 1961 recording. 
The solid line shows the
actual, performed tempo, while the dashed horizontal line is placed at
the indicated tempo of 132 b.p.m. The figure has three important
lessons: (1) observed speed varies around intended tempo; (2) 132 b.p.m.\ is
not necessarily the tempo a performer will choose despite the
indication; and (3) performers have other tempo intentions which are
not marked, like the pronounced slow-down in measures 7--8.

Estimating intended {\em tempi} would be reasonably simple, perhaps, 
if the locations of the tempo changes were known. In such a case,
the average of tempi between changes may be a good estimate as
could the slope of known speed-ups or slow-downs. However, performers
take liberties with these decisions, exactly the liberties we would
like to discover. This suggests employing a switching model with a
small number of discrete states.

We propose a Markov model for $S$ on four
states for four different performance behaviors
with transition probability
diagram given by \autoref{fig:transmat}.
\begin{figure}[tb!]
  \centering
  \tikzstyle{switch}=[rectangle,
  thick, minimum size=1cm, draw=black]
  \begin{tikzpicture}[>=latex,text height=1.5ex,text depth=0.25ex]
    \matrix[row sep=0.1cm,column sep=.25cm] {
      \node (S4) [switch] {$4$}; &&&&& & \node (S22) [switch] {$2$};\\
      &\node (S1) [switch] {$1$}; &&&& \node (S2) [switch] {$2$}; \\
      \\ \\ \\ \\ \\ \\
      &\node (S3) [switch] {$3$};\\
      \node (S33) [switch] {$3$};\\
    };
    \path[->]
    (S1) edge [bend left]  (S4)
    (S4) edge [bend left] (S1)
    (S1) edge [bend left] node [above] {$p_{12}$}(S22)
    (S22) edge (S2)
    (S2) edge [bend left] (S33)
    (S1) edge [bend right] node [left] {$p_{13}$}(S33)
    (S33) edge (S3)
    (S3) edge [bend right] node [right] {$p_{32}$}(S22)
    (S3) edge [loop left](S3)
    (S2) edge [loop above] node [left] {$p_{22}$}(S2)
    (S3) edge node [left] {$p_{31}$}(S1)
    (S2) edge node [above] {$p_{21}$}(S1);
   \path[->] (S1) edge [out=300,in=330,looseness=8] node [below right]
   {$p_{11}$} (S1);
  \end{tikzpicture}
  \caption{Transition diagram. The four states are: constant tempo
    (1), deceleration (2), acceleration (3), and emphasis (4).\label{fig:transmat}}
\end{figure}
The 4 switch states correspond to 4 different behaviors for the
performer: (1) constant tempo, (2) speeding up, (3) slowing down, and
(4) single note stress. As shown in the diagram, we allow only certain
transitions for musical reasons and for estimability. The marked
transition probabilities are sufficient to infer the remainder. One
can imagine that a performer will remain mainly in the state 1 with
departures to states 2 and 3 either due to markings by the composer,
or, absent these, for interpretive reasons referred to collectively as
{\em rubato}, which
translates literally as ``stolen time''. The fourth
state, stress, corresponds to {\em tenuto}, a common feature of
musical performance. These stresses may be marked with a line over the
note in question, but are more often a feature of performer taste,
corresponding to a longer-than-written duration for a particular
note. Such emphases occur for a variety of musical purposes---emphasis
of the beat in running notes, the top of a
phrase, a ``landing point'' where a phrase ends, etc.---but are always
within the frame of constant tempo. Thus we allow stress to occur only
after and before notes in state 1. Furthermore, we cannot allow
state 2 or state 3 to return immediately to state 1, or else ``stress'' could
happen through these pathways. We impose related constraints for a transition from state 2
to state 3 and vice versa. Essentially, transitions into these states must remain
there before leaving. Thus, the entire transition diagram is
fully determined. This process can  be viewed equivalently as a
second-order Markov chain.
% We discuss some potential improvements at the end
% of~\autoref{sec:analys-chop-mazurka}.

Generally, we feel that this model should be broadly applicable across
composers and time periods as well as instrumentation. That
is, it should work equally well for compositions by Mozart, Bach,
Beethoven, or Stravinsky. While some composers, especially those that
are more modern, have a more varied use of time signature and rhythm
than the music we examine here, these generally require even more stringent
adherence to ``steady tempo''. Some of Chopin's other compositions
present more severe departures from steady tempo (the Nocturnes,
for example), but we intend our model to be able to capture these
features through states 3 or 4. We return to this issue in~\autoref{sec:prior-sens-gener}.

Our data gives $y_i$ as the observed tempo (in b.p.m.) of the note (or
chord) of the $i^{th}$ note onset in Chopin's Mazurka Op.\ 68 No.\ 3. The
hidden continuous variable ($X_i$) is 
taken to be a two component vector with the first component being the
prevailing tempo and the second the amount of acceleration. The amount, or
existence, of acceleration is determined by the current and previous
switch states. We use $\ell_i$ to denote the musical duration of
a particular note as given by the written score. Because, in this
piece, each measure contains three quarter-notes, (\quarternote), a quarter-note has $\ell_i=1/3$, an
eighth note (\eighthnote) has $\ell_i=1/6$, etc. In more complicated
music with changing time signatures or instances where the notation
doesn't necessarily correspond with the time signature, more care
would be required. The observed tempo is already normalized to account
for variable note durations, but the intentional tempo and its
variance should be proportional to $\ell_i$. When the performer is in state 1 (or
transits in and out of state 4), we take the prevailing tempo as
constant with no acceleration: $X_{i+1} = X_i$. 
Corresponding to these configurations, the parameter
matrices are given in \autoref{tab:parmats} (transition equation) and
\autoref{tab:parmats2} (measurement equation).
\begin{table}
  \caption{Parameter matrices of the transition equation for the switching state space model.\label{tab:parmats}}
\centering
\begin{tabular}[h!]{@{}cccccc@{}}
\toprule
%&&&\multicolumn{3}{c}{Transition equation}\\
  \multicolumn{2}{c}{Switch states} &\phantom{a}& \multicolumn{3}{c}{parameter
                                      matrices}\\
  \cmidrule{1-2} \cmidrule{4-6}
  $s_i$ & $s_{i-1}$ && $d$ & $T$ & $Q$ \\
  \midrule
  $1$ &  $1$ && 0 & $\begin{pmatrix}1&0\\0&0\end{pmatrix}$ 
                 & $\begin{pmatrix}0&0\\0&0\end{pmatrix}$\\
  $2$ & $1$ && $\begin{pmatrix} \ell_i\mu_{\textrm{acc}}\\ \mu_{\textrm{acc}}\end{pmatrix}$ 
                                    & $\begin{pmatrix} 1 & 0 \\ 0 &
                                      0 \end{pmatrix}$ 
   & $\sigma_{\textrm{acc}}^2\begin{pmatrix} \ell_i^2 & \ell_i\\ \ell_i & 1 \end{pmatrix}$\\
  $3$ & $1$ && $\begin{pmatrix} -\ell_i\mu_{\textrm{acc}}\\ -\mu_{\textrm{acc}}\end{pmatrix}$ 
                                    & $\begin{pmatrix} 1 & 0 \\ 0 &
                                      0 \end{pmatrix}$ 
        & $\sigma_{\textrm{acc}}^2\begin{pmatrix} \ell_i^2 & \ell_i\\ \ell_i & 1 \end{pmatrix}$\\
  $4$ & $1$ && $\begin{pmatrix}0\\\mu_{\textrm{stress}}\end{pmatrix}$ 
                                     & $\begin{pmatrix}1&0\\0&0\end{pmatrix}$
          & $\begin{pmatrix}0&0\\0&\sigma_{\textrm{stress}}^2\end{pmatrix}$\\
  $2$ & $2$ && 0 & $\begin{pmatrix} 1 & \ell_i \\ 0 & 1 \end{pmatrix}$ 
        & $\begin{pmatrix}0&0\\0&0\end{pmatrix}$\\
  $3$ & $2$ && $\begin{pmatrix} -\ell_i\mu_{\textrm{acc}}\\ -\mu_{\textrm{acc}}\end{pmatrix}$ 
                                    & $\begin{pmatrix} 1 & 0 \\ 0 &
                                      0 \end{pmatrix}$ 
        & $\sigma_{\textrm{acc}}^2\begin{pmatrix} \ell_i^2 & \ell_i\\ \ell_i & 1 \end{pmatrix}$\\
  $1$ & $2$ && $\begin{pmatrix} \mu_{\textrm{tempo}}\\0\end{pmatrix}$ & 0
        & $\begin{pmatrix} \sigma^2_{\textrm{tempo}} & 0\\ 0 & 0 \end{pmatrix}$\\
  $3$ & $3$ && 0& $\begin{pmatrix} 1 & \ell_i \\ 0 & 1 \end{pmatrix}$ 
        & $\begin{pmatrix}0&0\\0&0\end{pmatrix}$\\
$2$ & $3$ && $\begin{pmatrix} \ell_i\mu_{\textrm{acc}}\\ \mu_{\textrm{acc}}\end{pmatrix}$ 
                                    & $\begin{pmatrix} 1 & 0 \\ 0 &
                                      0 \end{pmatrix}$ 
        & $\sigma_{\textrm{acc}}^2\begin{pmatrix} \ell_i^2 & \ell_i\\ \ell_i & 1 \end{pmatrix}$\\
  $1$ & $3$ && $\begin{pmatrix} \mu_{\textrm{tempo}}\\0\end{pmatrix}$ & 0
          & $\begin{pmatrix} \sigma^2_{\textrm{tempo}} & 0\\ 0 & 0 \end{pmatrix}$\\
  $1$ &  $4$ && 0 & $\begin{pmatrix}1&0\\0&0\end{pmatrix}$ 
        & $\begin{pmatrix}0&0\\0&0\end{pmatrix}$\\
  \bottomrule
\end{tabular}
\end{table}
\begin{table}
  \caption{Parameter matrices of the measurement equation for the switching state space model.\label{tab:parmats2}}
\centering
\begin{tabular}[b!]{@{}ccccc@{}}
\toprule
%&&&\multicolumn{3}{c}{Measurement equation}\\
  Switch states &\phantom{a}& \multicolumn{3}{c}{parameter
                              matrices}\\
  \cmidrule{1-1} \cmidrule{3-5}
  $s_i$ && $c$ & $Z$ & $G$\\
  \midrule
  $4$  && 0 & $\begin{pmatrix} 1 & 1 \end{pmatrix}$ &
                                                                  $\sigma^2_\epsilon$\\
  else && 0 & $\begin{pmatrix} 1 & 0 \end{pmatrix}$ &
                                                                  $\sigma^2_\epsilon$\\
\bottomrule
\end{tabular}
\end{table}
So for any performance, we wish to estimate
the following parameters: $\sigma_{\textrm{tempo}}^2$, $\sigma_{\textrm{acc}}^2$, $\sigma^2_{\textrm{stress}}$,
$\sigma_\epsilon^2$, the probabilities of the transition matrix (there
are 7), and means $\mu_{\textrm{tempo}}$, $\mu_{\textrm{acc}}$, and $\mu_{\textrm{stress}}$. Lastly, we have the initial state distribution
\[
x_1\sim N\left( \begin{pmatrix}\mu_1\\0\end{pmatrix}
  ,\ \begin{pmatrix} \sigma^2_1 & 0\\0 & 0
  \end{pmatrix}\right)\; \; \textrm{where} \; \; s_1=1.
\]

To clarify this model, we explicate two different behaviors: discrete
sequence $1\rightarrow 4\rightarrow 1$ (emphasis within constant tempo) and discrete sequence
$1\rightarrow 1\rightarrow 2$ (constant tempo to slowing down). In the
first case, the state space system has the following configurations

% {\footnotesize
\begin{equation*}
  \begin{aligned} 1\rightarrow 4 && 4\rightarrow 1\\ x_{2}
    &= \begin{pmatrix} 0\\ \mu_{\textrm{stress}} \end{pmatrix}
    + \begin{pmatrix}1&0\\0&0\end{pmatrix} x_{1} +
    \mbox{N}\left(0,\ \begin{pmatrix}0&0\\0&\sigma_{\textrm{stress}}^2\end{pmatrix}\right)
    & x_{3} &=
    \begin{pmatrix}1&0\\0&0\end{pmatrix} x_{2} \\ y_2 &= (1\quad 1) x_2
    + \mbox{N}(0,\ \sigma_\epsilon^2) & y_3 &= (1\quad 0) x_3 +
    \mbox{N}(0,\ \sigma_\epsilon^2),
  \end{aligned}
\end{equation*}
% }%
while in the second
% {\footnotesize
\begin{equation*}
  \begin{aligned} 1\rightarrow 1 && 1\rightarrow 2\\ x_{2} &=
    \begin{pmatrix}1&0\\0&0\end{pmatrix} x_{1} & x_{3}
    &= \begin{pmatrix} \ell_i\mu_{\textrm{acc}}\\
      \mu_{\textrm{acc}}\end{pmatrix} +
    \begin{pmatrix}1&0\\0&0\end{pmatrix} x_{1} + \mbox{N}\left(0,\
      \sigma_{\textrm{acc}}^2\begin{pmatrix} \ell_i^2 & \ell_i\\ \ell_i &
        1 \end{pmatrix}\right)\\ y_2 &= (1\quad 0) x_2 + \mbox{N}(0,\
    \sigma_\epsilon^2) & y_3 &= (1\quad 0) x_3 + \mbox{N}(0,\
    \sigma_\epsilon^2).
  \end{aligned}
\end{equation*}
%}%
Recall that in any case $y_i$ is a scalar and $x_i \in \mathbb{R}^2$.

\subsection{Estimation and computational issues}
\label{sec:computational-issues}

To understand the performance decisions of individual musicians, we
wish to simultaneously estimate $\theta$, $S$, and $X$. Because the
switch states $S$ and the continuous states $X$ are both hidden, this becomes
an NP-hard problem. In particular, there are approximately $4^n$ possible paths
through the switch variables, so evaluating the likelihood to maximize
over $\theta$ via the Kalman filter at each path is intractable. 
\citet{GhahramaniHinton2000} give a variational approximation to
estimate $\theta$ without also estimating $S$, but, as our goal is to
learn both, we use the particle filtering approximation described by
\citet{FearnheadClifford2003}. \cite{WhiteleyAndrieu2010} refer to
this algorithm as the Discrete Particle Filter, and it can be seen as
an instance of the ``Beam Search'' optimization
technique~\citep{Bisiani1992}. The details are given in
\autoref{alg:dpf} but the intuition is as follows: (1) for the first
few time points, evaluate
one step of the Kalman filter for each possible subsequent discrete
state and store all these values; (2) calculate weights for each path
by updating previous weights with the likelihood multiplied by the transition probability;
(3) continue through time until the number of stored values exceeds
some threshold storage limit; (4) from that point forward, subselect
the ``best'' paths using a sampling scheme.
\begin{algorithm}[t!]
  \caption{Discrete particle filter\label{alg:dpf}}
  \begin{algorithmic}[1]
  \STATE {\bfseries Input:}
  $Y$, $\theta$, $\pi_1$ probability vector over initial states
  (paths), $B$ beam width
  \FOR{$i=1$ {\bfseries to} $n$}
  \STATE Set $b_i=|\{\pi_i>0\}|$, the number of current paths
  \STATE Use %\autoref{alg:kalman}
  the Kalman filter to calculate the 1-step likelihood
  $\mathcal{L}_i$ for each path and every potential state $s_{i+1}$ resulting in $b_i|S|$ particles
  \STATE Set $\pi_{i+1} \leftarrow \pi_i\mathcal{L}_i p_i$: multiply the path
  probability by the likelihood and the probability of
  transitioning. Normalize $\pi$.
  \STATE Set $b_{i+1}=|\{\pi_{i+1}>0\}|$ . If $b_{i+1} > B$, resample the
  weights to get $B$ non-zero weights and renormalize
  \ENDFOR
  \STATE Return $B$ paths $\{S_b\}_{b=1}^B$ along with their weights $\pi_{n}$.
\end{algorithmic}
\end{algorithm}
These paths can be
selected greedily, retaining only the highest values to that point,
though we use the resampling procedure of
\citet{FearnheadClifford2003} which is designed to 
approximate the full discrete distribution over paths with a subset
of support points by minimizing the mean squared
error.

\autoref{alg:dpf} returns $B$ paths through the discrete states along with their weights for a 
particular parameter value $\theta$. One
can view this as a (approximate) distribution over paths conditional
on $\theta$. Instead, we will simply take the path with the highest
weight for inference via penalized maximum likelihood. Thus, the
likelihood of a particular parameter vector $\theta$ is evaluated by
computing the best path with \autoref{alg:dpf} and then using that best
path with the Kalman filter.%\autoref{alg:kalman}.

\subsection{Penalized maximum likelihood}
\label{sec:penal-maxim-likel}

Even without the latent discrete process, parameter estimation in
state-space models is a difficult problem, often plagued by spurious
local minima and non-identifiability. The addition of discrete states
only exacerbates this issue. However, for the present application, we
have reasonable informative prior information for many of the
parameters. The three mean parameters $\mu_{\textrm{tempo}}$,
$\mu_{\textrm{acc}}$ and $\mu_{\textrm{stress}}$ have sign
restrictions in addition to reasonable constraints their magnitude:
average tempo should be around the indicated 132 b.p.m., the average
amount of acceleration should probably be less than the size of a
stress. We also have can make musically informed choices about the
probabilities of transitioning between states: self-transitions should
be reasonably likely, long periods of speeding up are less likely than
long periods of slowing down which are less likely than long periods
in the constant tempo state.

Because of this information, we use
informative priors as penalties on all the parameters we
estimate. This has the effect of introducing extra curvature to the
optimization problem as well as conforming with musical intuition. The
specific choices are shown in \autoref{tab:priors}. 
\begin{table}[t]
    \caption{Informative prior distributions for the music model}
  \label{tab:priors}
  \centering
  \begin{tabular}{@{}rcll@{}}
    \toprule
    Parameter & \phantom{a} & Distribution & Prior mean\\
    \midrule
    $\sigma^2_{\epsilon}$ & $\sim$ & Gamma$(40,\ 10)$ & 400 b.p.m.$^2$\\
    $\mu_{\textrm{tempo}}$ & $\sim$ & Gamma$(\overline{Y}^2/100,\ 100
                                      /\overline{Y})$ & $\overline{Y}$
                                                        b.p.m.\\
    $-\mu_{\textrm{acc}} $ & $\sim$ & Gamma$(15,\ 2/3)$ & 10 b.p.m.\\
    $-\mu_{\textrm{stress}} $ & $\sim$ & Gamma$(20,\ 2)$ & 40 b.p.m.\\
    $\sigma^2_{\textrm{tempo}} $ & $\sim$ & Gamma$(40,\ 10)$ & 400
                                                               b.p.m.$^2$\\
    $\sigma^2_{\textrm{acc}} $ & $=$ & 1 & 1 b.p.m.$^2$\\
    $\sigma^2_{\textrm{stress}} $ & $=$ & 1 & 1 b.p.m.$^2$\\
    $p_{1,\cdot}$ & $\sim$ & Dirichlet$(85,\ 5,\ 2,\ 8)$ \\
    $p_{2,\cdot}$ & $\sim$ & Dirichlet$(4,\ 10,\ 1,\ 0)$ \\
    $p_{3,\cdot}$ & $\sim$ & Dirichlet$(5,\ 3,\ 7,\ 0)$ \\
    \bottomrule
  \end{tabular}
\end{table}
We fix $\sigma^2_{\textrm{acc}}$ and $\sigma^2_{\textrm{stress}}$ to
be 1 after numerical experiments suggested
that they were poorly identified. Essentially, large values of these
variances make the stress and deceleration states difficult to
separate, so other values of similar magnitude make little
difference. Given the use of priors, one could interpret our
procedure from a Bayesian perspective. However, we do not estimate full
posteriors, only the maximum \emph{a posteriori} value, so our prior
specification choices
are not intended to reflect tail uncertainty. We defer justification and
discussion of these choices to 
\autoref{sec:prior-sens-gener}. 

\subsection{Is this model reasonable}
\label{sec:this-model-reas}

It is reasonable to ask whether a simple model such as this can
accurately represent performance practice without removing musically important
information. In a statistical sense, this question is similar to the
problem of tuning parameter selection in nonparametric
estimation. Specifically, we do not want this model to
``over-smooth'' the performance, eliminating information necessary for
listener appreciation. One way to examine such a question 
is to generate a performance using the smoothed tempos resulting from
the fitted model and compare it aurally with the original recording.
\citet{GuRaphael2012} evaluate this question empirically: they surveyed nine
graduate piano majors at a 
major conservatory on twelve different piano excerpts, both performed
and synthesized. The pianists
were not meaningfully able to distinguish between the two in the
majority of experiments. We expect that a similar
study with our model would yield better results. In the Supplementary
Material~\citep{McDonaldMcBride2021a}, we allow the reader to decide for
themselves: we have 
included a MIDI recording derived from Sviatoslav Richter's 1976
recording as well as one synthesized using our model. The only
difference is the tempos of the individual beats.

The generative model in \citet{GuRaphael2012} is also a switching
model on four states like ours. It is quite a bit simpler, however, in terms of
the transition matrix depicted in \autoref{fig:transmat}. It is only first-order, so states 2 and 3 can
be entered and left immediately. There's also no ability to go from
state 3 to 2 or 2 to 1. This precludes the common feature of
slowing down at the end of a phrase before returning to a new tempo. Furthermore, and importantly, their parameters are not
estimated from data but chosen by eye. Comparing our model to theirs, we found that
ours is more robust in that it is less likely to make spurious
excursions to states 2 and 3 and more accurately uses state 4. It also
has significantly lower RMSE on the data despite having only two
additional parameters. A more careful comparison with this model is given in the
Supplementary Material~\citep{McDonaldMcBride2021a}.

While an additive state space model is relatively easy to understand,
some music theorists \citep[for example]{Mead2007} have argued that
musicians make multiplicative tempo adjustments. That is, the ratio
between the tempo of the current note and that of the previous note is
important rather than their difference. Such a conception is
fundamental to musical notation (quarter notes, eighth notes, etc) and
frequently used to specify tempo changes within a piece of
music. Unfortunately, switching linear models are challenging to
estimate and non-linear models are only more so. We examine a
multiplicative model in the Supplementary Material. This model produces
very reasonable interpretations of individual performances, but
unfortunately, it is less
useful for comparing performances.

\section{Analysis of Chopin's Mazurka Op.\ 68 No.\ 3}
\label{sec:analys-chop-mazurka}

We use the model and procedures developed above to estimate the
parameters and performance choices for all 46 recordings of Chopin's
Mazurka. Here we describe the inferences our model allows on some
representative performances, describe performance groupings based on
the estimated parameters, contrast our model with some alternative
approaches to smoothing, and discuss some difficulties we encountered. All simulations and empirical calculations were performed with
\texttt{R}~\citep{R-Core-Team2019} and C++ via \texttt{Rcpp}~\citep{Eddelbuettel2013}. Figures and tables are generated
using the \texttt{tidyverse} family of
packages~\citep{Wickham2017, Wickham2016}.
% Dendrograms combined with
% heatmaps for the proximity matrices were created with the
% \texttt{heatmaply} package~\citep{GaliliOCallaghan2017}.
% The Supplemental Materials were created
% with \texttt{knitr} and
% \texttt{rmarkdown}~\citep{Xie2019,XieAllaire2018,Xie2015}.
Most
computations were implemented in parallel on a
% the Carbonate\footnote{This research was supported in part by Lilly
%   Endowment, Inc., through its support for the Indiana University
%   Pervasive Technology Institute.}
large
memory computer cluster via
the \texttt{batchtools} package~\citep{LangBischl2017}. 

\subsection{Musical analysis}
\label{sec:musical-analysis}

Throughout his life, Fr\'ed\'eric Chopin composed dozens of Mazurkas,
of which 58 have been published. Inspired by a traditional Polish
dance, these pieces gave Chopin an idiomatic style upon which to
elaborate a wide variety of different compositional techniques, a
practice German and Italian composers had employed frequently over the previous 3
centuries~\citep{BurkholderGrout2014}. Repetition of themes, figures, or even small motives plays
a central role in both the traditional dance and Chopin's compositions
as do particular rhythmic gestures~\citep{Kallberg1996}, especially the
dotted-eighth sixteenth note pattern on the first beat of a measure. 

Chopin's Op.\ 68 Mazurkas are a set of four similar works, published
posthumously in 1855. The Op.\ 68 No.\ 3, which we analyze here, was
composed in 1830, when Chopin was 20 years old. Around this time,
Chopin, already a piano virtuoso and accomplished composer, left his
native Warsaw and settled in Paris, where we would remain until his
death in 1849.

This Mazurka has a rather simplistic ternary structure with two outer
sections and a contrasting middle (ABA). The first A section is made
up of four eight-bar phrases ($aaba$). The first phrase is echoed by the
second phrase: they are nearly identical, with the two exceptions
being that (1) the
second is marked {\em piano} (soft) rather {\em forte} (strong) and (2)
the second ends on the tonic (F major) rather than the dominant
(C major). The fourth eight-bar phrase is an exact repetition of the
second. The second A section is a repeat of the first two
eight-bar phrases of the beginning. The intervening B section is 12
bars long, divided into three four-bar groups. The first four bars are
simply a repeated interval of a perfect $5^{th}$ in the left
hand. This {\em ostinato} will continue for the whole section. The remaining
eight measures consist of a four-bar phrase in the right hand repeated twice. The second differs from the first only on the final two notes, preparing the
recapitulation of the A section.

In terms of tempi, the B section is indicated to be faster, with the
marking {\em Poco pi\`u vivo} (a little livelier). The B section ends
with a {\em ritardando} into the following A section. The $b$ section ends
with a {\em fermata} in measure 24, indicating an arbitrary
elongation while the piece
concludes with a two-measure long {\em ritardando}. Throughout, 
frequent markings prescribe emphasis of the third beat of each measure. This
emphasis is in keeping with the mazurka style, an intentional
thwarting of the listener's expectation of first-beat emphasis.

\begin{figure}[t]
  \centering
  \includegraphics[width=.9\textwidth]{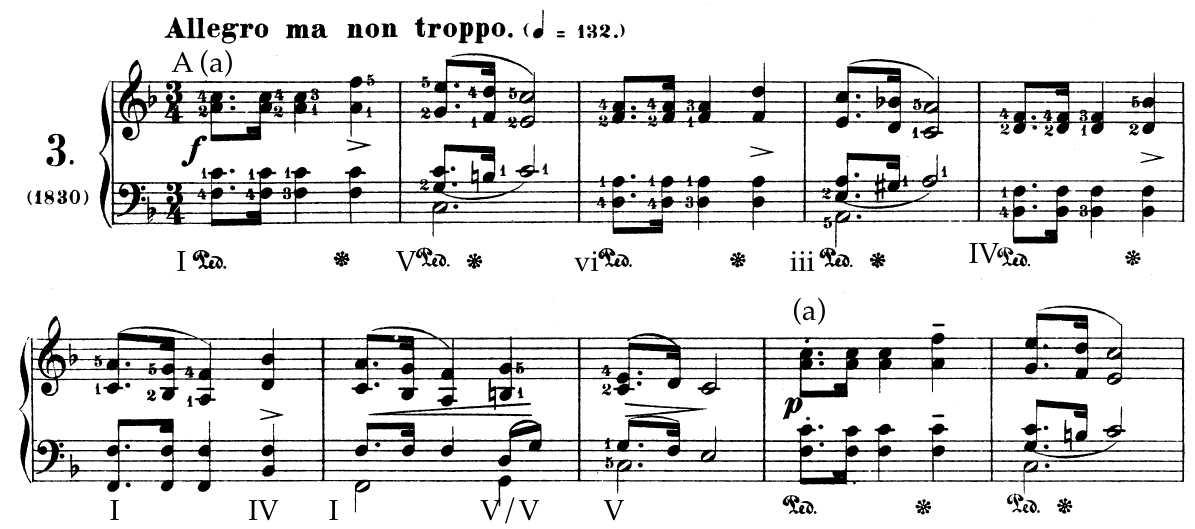}
  \caption{The first ten measures of Chopin's Mazurka Op.\ 68, No.\
    3. The harmonic progression is indicated below the staff in Roman
    numerals. Sections are marked above the staff, e.g., A
    (a). Analysis by the authors. This image comes from the complete
    score published by Bote and Bock in 1880. This composition is in
    the public domain, and the score is freely available via the
    International Music Score Library Project.}
  \label{fig:mazurka-10}
\end{figure}

\autoref{fig:mazurka-10} shows the first ten measures of the musical
score with annotations for the sections discussed above and the
harmonic progression in Roman numerals below the staff. 
The harmonies are standard, in fact, they are essentially the same as
those of Pachelbel's {\em Canon}, familiar to many as ``that song
played at weddings.'' These harmonies, combined with the
rhythmic repetition suggests a further division of this and
all analogous sections into three small groupings: two two-measure
phrases, followed by a four-measure phrase.

As a performer, these harmonic, rhythmic, and structural analyses aid
in interpretation. The performer needs to decide how to emphasize or
deemphasize these demarcations with slight or overt tempo or dynamic
alterations. In a live performance, she could use physical motion to
further suggest a particular interpretation. She can choose to
emphasize long phrases, in this 
case, phrases of eight measures, or the shorter sub-phrases. Because
of the repetition of similar phrases, she may choose to emphasize the
long phrase on the first occurrence and shorter sub-phrases later on
for variety, for example. While the musical structure suggests such
possible interpretations, the performer must make these choices on
her own, and may even alter those decisions from performance to
performance.

\subsection{Archetypal performances}
\label{sec:arch-perf}

\begin{figure}[t]
  \centering
  \includegraphics[width=.9\textwidth]{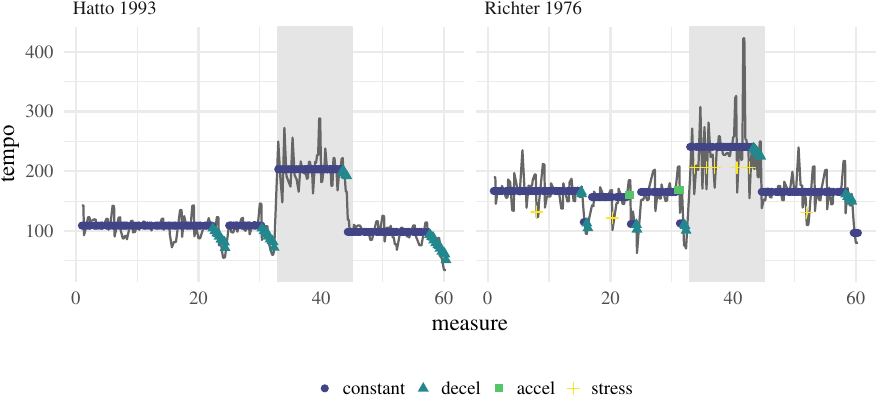}
  \caption{Inferred performer choices for two recordings. }
  \label{fig:archetypal}
\end{figure}
Here we will carefully investigate the interpretive
decisions implied by our estimated model for three rather different performances. \autoref{fig:archetypal} shows the inferred state sequence for
recordings made by Joyce Hatto in 1993 and Sviatoslav Richter in
1976. The B section is
shaded in gray to better illustrate the formal divisions discussed above.

Our estimated model suggests that these two performers are quite different
from each other. Hatto maintains a constant tempo carefully, remaining
in state 1 with the exception of four periods of deceleration. All
four periods coincide with the most significant phrase endings: at the
end of the A section at measure 32, the end of the B section at
measure 48, at the end of the piece, and the minor transition from
$b\rightarrow a$ in the first A section (measure 24). According
to our inferred model, she never accelerates or uses the transitory
stress state.

In contrast, Richter uses all four states from our model. The short
blips of acceleration before the B section and before the
$b \rightarrow a$ transition are slightly out of place, and are
likely better labelled as ``constant'', but these state transitions
describe more severe decelerations than the model's linear
assumption would allow (the multiplicative model in the Supplementary
Material is much better). Richter uses stress frequently. Some may well be
attributable to larger variance around constant tempo (picked up as
frequent stress rather than larger $\sigma^2_\epsilon$), but most
correspond to interesting note emphases, for example the second beat
of measure 20. This note is essentially a minor phrase ending, but it is
also marked in the score with a {\em sforzando} (with sudden
emphasis). It's the first of two such occurrences in the piece, the
second coming four measures later on the {\em fermata}, Richter's slowest
note in the entire piece. Richter likely chooses to make this
prescribed emphasis with a sudden slow down in part because it takes
place within the context of an already loud passage, precluding the
use of extra volume.
\begin{table}[tb]
  \centering
  \caption{The estimated parameters for performances by Richter and
    Hatto.}
  \resizebox{\linewidth}{!}{%
  \begin{tabular}{@{}lrrrrrrrrrrrr@{}}
    \toprule
    & $\sigma^2_\epsilon$ & $\mu_{\textrm{tempo}}$
    & $\mu_{\textrm{acc}}$ & $\mu_{\textrm{stress}}$
    & $\sigma^2_{\textrm{tempo}}$ & $p_{11}$ & $p_{12}$ & $p_{22}$
    & $p_{31}$ & $p_{13}$ & $p_{21}$ & $p_{32}$\\
    \midrule
    Richter 1976 & 426.70 & 136.33 & -11.84 & -34.82 & 439.38
                                  & 0.85 & 0.05 & 0.74 & 0.44 & 0.02 & 0.25 & 0.17\\
    Hatto 1993 & 405.57 & 130.36 & -13.57 & -27.93 & 408.99
                                  & 0.94 & 0.03 & 0.82 & 0.36 & 0.01 &
                                                                       0.16 & 0.19\\
    Cortot 1951 & 403.71 & 182.84 & -21.43 & -45.67
    & 460.82 & 0.92 & 0.02 & 0.71 & 0.34 & 0.03 & 0.23 & 0.09\\
    \bottomrule
  \end{tabular}
}
\label{tab:two-perf-parm}
\end{table}
\autoref{tab:two-perf-parm} shows the estimated parameters for these
two performances.\footnote{In the Supplementary Material, we provide
  the estimates along with some measures of uncertainty for all 46
  recordings.} Richter has larger observation variance,
$\sigma^2_{\epsilon}$, slightly faster average tempo, lower
acceleration, and larger stress. He also has a larger tempo variance,
meaning that returns to state 1 can start at relatively different
tempos. On the other hand, Hatto is much more
likely to remain in states 1 or 2. These inferences are largely
consistent with the visual messages of \autoref{fig:archetypal}. The
variability definitely increases around the constant tempo in
Richter's performance and he uses faster overall tempos in both the A and
B sections.
While these two performances are quite different from each other, they
also display similarities. Both take a faster tempo in the B
section versus the A sections. Both performers slow down at the end of
the piece, at the end of the B section, immediately preceding the B section, and at the
$b\rightarrow a$ transition.

\begin{figure}[t]
  \centering
    \includegraphics[width=.9\textwidth]{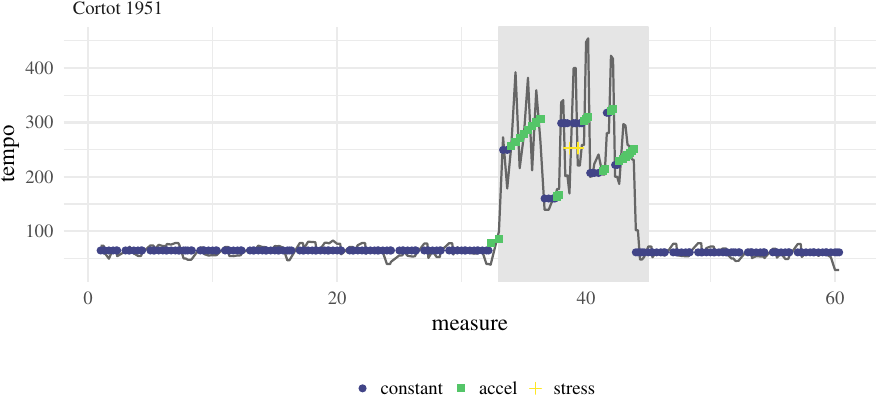}
  \caption{Inferred performance choices for Alfred Cortot's 1951
    recording.}
  \label{fig:cortot}
\end{figure}

Alfred Cortot's 1951 performance is displayed
in~\autoref{fig:cortot}. Both in terms of the parametric model we
propose, and if we simply compare the vectors of note-by-note tempos
(discussed in more detail below), this performance is an
outlier. Cortot never uses the deceleration state, and he remains in
constant tempo for the entirety of both A sections. While the model
describes his performance well, it also illustrates a deficiency of
this approach: Cortot, more than any other performer, has large
contrasts between the A and B sections. His A section is the slowest
of all 46 recordings at around 64 b.p.m., half the marked
tempo. The next slowest is Maryla Jonas's recording at around 84
b.p.m. Meanwhile, his B section is among the fastest of all the
recordings and contains the fastest individual note. Additionally,
there is stunningly little tempo variability in his A sections, but
dramatic variation in the B section coupled with frequent uses of the
acceleration and emphasis states. Taken together, Cortot's performance
may be better described by estimating our model separately on the two
sections.\footnote{\citet{Cook2013} suggests that this recording is not
  due to Cortot at all but part of a scandal at the Concert Artist label
  referred to as the ``Hatto hoax'', wherein her husband, owner of the
  label, released over 100 recordings made by others but listing her
  as the performer.}

\subsection{Comparing performances}
\label{sec:clust-music-perf}

To better understand how the 46 recordings relate to each other,
we measured the distance between their vectors of paramater
estimates. Because the parameters have 
different scales, have different domains, and can covary, we
use Mahalanobis distance to scale by the inverse of the prior
covariance. That is, the distance between performance $i$ and
performance $j$ is given by
\begin{equation}
  \label{eq:parameter-distance}
d_{i,j} = \left(\hat\theta_i-\hat\theta_j\right)^\top\Omega\left(\hat\theta_i-\hat\theta_j\right)
\end{equation}
where $\hat\theta$ corresponds to the estimates in,
e.g.~\autoref{tab:two-perf-parm}, and the prior precision, $\Omega=\Sigma^{-1}$,
is calculated  based on
the distributions in \autoref{tab:priors}.
We use the prior covariance rather than the covariance
of the estimated parameters because any parameters that are poorly
identified by the model will have small estimated variance, and
therefore dominate the distance calculation. Using the prior avoids
this pathology and while still properly accounting for scale and
structural dependence.
% The mean and
% variance parameters, $\sigma^2_\epsilon$, $\mu_{\textrm{tempo}}$,
% $\mu_{\textrm{acc}}$, $\mu_{\textrm{stress}}$, and
% $\sigma^2_{\textrm{tempo}}$, are straightforward since they are
% independent under the prior. In the cases of the probabilities, we use
% the covariance of the Dirichlet distribution
% \begin{equation}
% \Sigma_{ij} :=
%   \alpha_0^{-2}(\alpha_0+1)^{-1}\left(\alpha_i\alpha_0 \delta_{ij}
%     - \alpha_i\alpha_j\right),
% \end{equation}
% with
% $\alpha_0=\sum_i\alpha_i$ and $\delta_{ij}$ the indicator that
% $i=j$.
% We then standardize each individual distance matrix to have a
% maximum distance of 1 and add them together so that the maximum distance
% between performances is 8.

\begin{figure}[t]
  \centering
  \includegraphics[width=.65\linewidth]{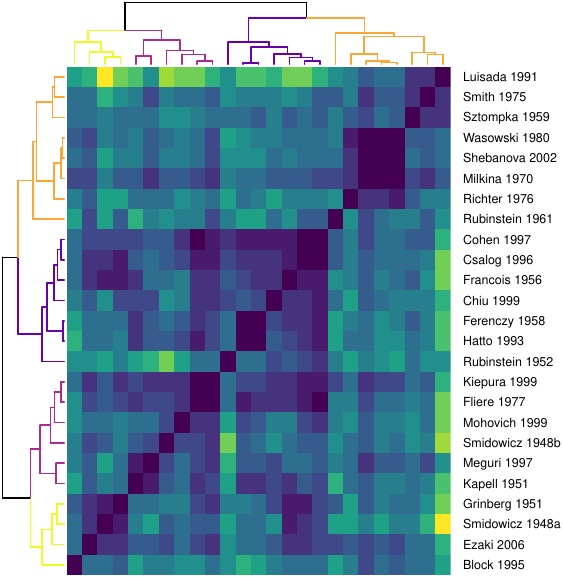}
  \caption{Distance matrix using the estimated parameters for all 46 performances. }
  \label{fig:dmats}
\end{figure}
\autoref{fig:dmats} shows the distance matrix calculated from the
estimated parameters for all 46 performances. The dendrogram helps to
visualize similarities between the performances, but should be taken
only as a heuristic. Across a variety of clustering procedures,
methods for choosing the number of clusters
\citep[e.g.,][]{TibshiraniWalther2001,DudoitFridlyand2002}
consistently suggest that there is only one cluster. The procedure
developed by \citet{TibshiraniWalther2001} chooses the number of
clusters by finding the first maximum of the Gap statistic, subject to
a measure of uncertainty. For the Chopin performances, the global maximum
occurs at 7 clusters, which is both implausible and not robust to
uncertainty. Nonetheless, for the purposes of organizing this section,
we will consider those performances within a ``group''  to be
more similar than across groups.

\begin{figure}[t]
  \centering
  \includegraphics[height=5in]{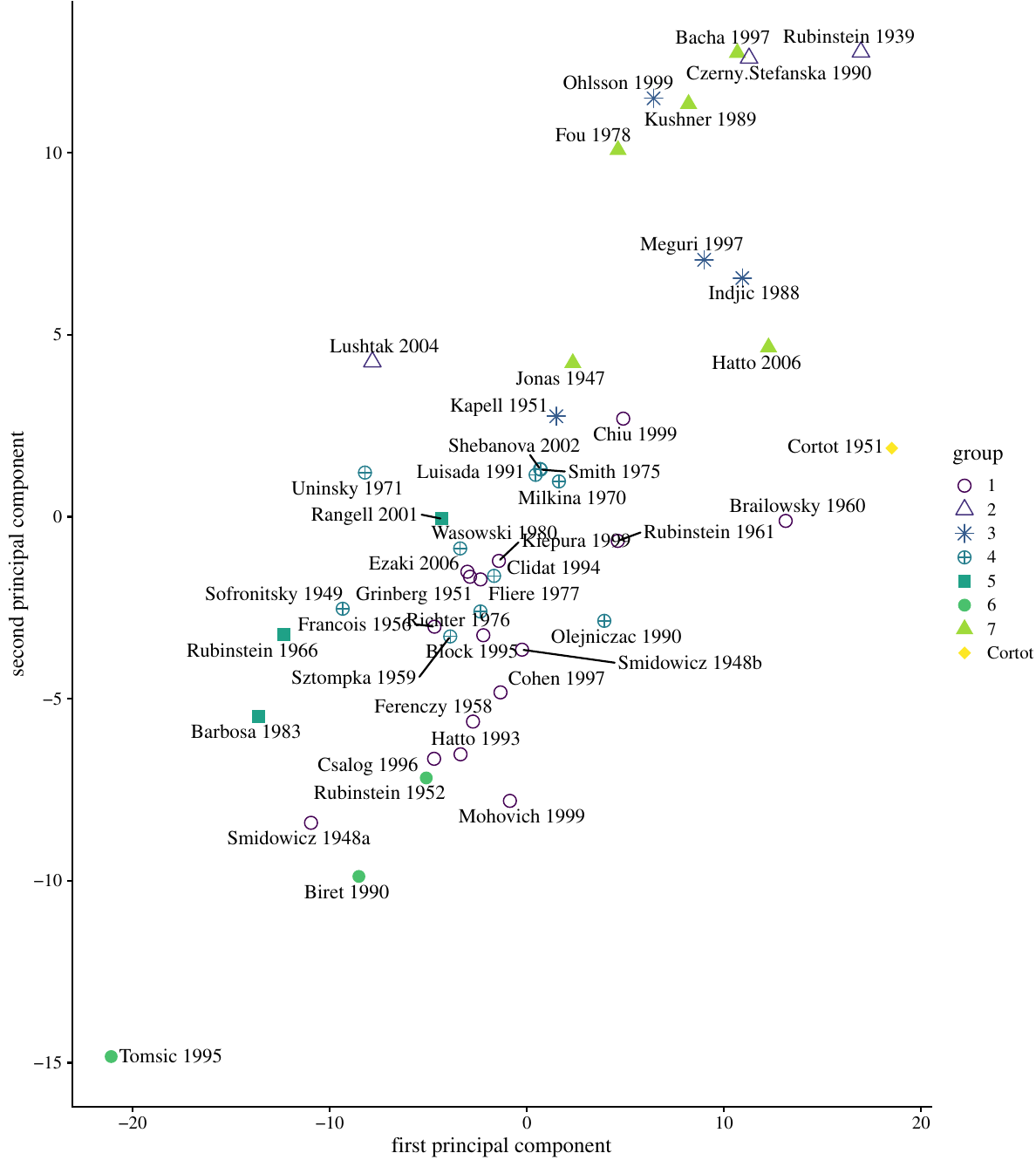}
  \caption{The first two principal components of the matrix of
    estimated parameters. Similar performances are indicated by
    shape.}
  \label{fig:pca}
\end{figure}

In order to inspect these performances visually, we follow the advice
of an anonymous reviewer and perform principal components analysis on
the matrix of estimated parameters. Only about 45\% of the variance is
explained by the first 2 components, and we would need 7 to explain
90\%, but \autoref{fig:pca} nonetheless corresponds somewhat closely
to the groups suggested by the dendrogram in \autoref{fig:dmats}.
In the Supplementary Material, we plot all the
inferred performance decisions by group and give the
factor loadings for the first few principal components. In the
remainder of this section, we describe
typical behaviors of the performances within a few groups that have relatively
small 
within-group variability.

The first group (indicated as $\circ$ in \autoref{fig:pca}) corresponds to
reasonably staid performances. This group is the largest and
corresponds to the block from Cohen to Brailowsky in
\autoref{fig:dmats}. In this group, the emphasis state is rarely visited
with the performer tending to stay in the constant tempo state with
periods of slowing down at the ends of phrases. Acceleration is almost never used. Furthermore, these
performances have relatively slow average tempos, and not much
difference between the A and B sections. Joyce Hatto's recording
in \autoref{fig:archetypal} is typical of this group.

\begin{figure}[t]
  \centering
  \includegraphics[width=.9\linewidth]{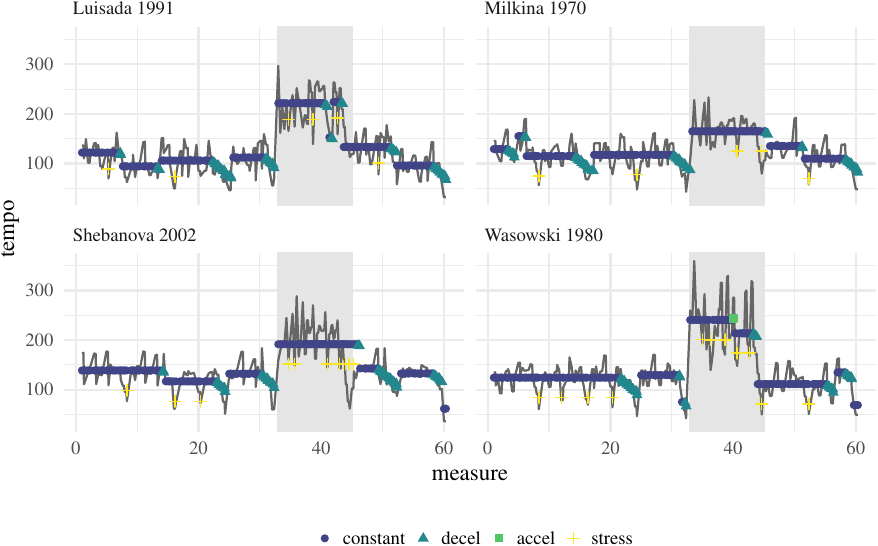}
  \caption{Four similar performances, all in the fourth group ($\oplus$).}
  \label{fig:similar}
\end{figure}

Recordings in the fourth group ($\oplus$ in \autoref{fig:pca}) are
those in the upper right of \autoref{fig:dmats}, from Olejniczac to
Richter. These recordings
tend to transition quickly between states, especially constant
tempo and slowing down, accompanied by frequent transitory
emphases. The probability of remaining in state 1 is the lowest while
the probability of entering state 2 from state 1 is 
the highest. The acceleration state is rarely visited. Four of
the most similar performances are in this group, shown in
\autoref{fig:similar}, along with Richter's 1976 recording.

The three performances in group six ($\bullet$) are actually quite like others,
but with small exceptions. Biret's 1990 performance is very much like
those in group 1, but with a much larger contrast between tempos in
the A and B sections. The recording by Rubinstein in 1952 is similar,
though with a faster A section that has less contrast with the B
section. Tomsic's 1995 performance is actually most
similar to those in group three ($\mathrlap{+}\times$), but played much faster and
with a large $\sigma^2_\epsilon$.

% \begin{figure}[t]
%   \centering
%   \includegraphics[width=.45\linewidth]{raw-data-clusters-1}
%   \caption{Distance matrix and dendrogram calculated using the
%     note-by-note tempo vector for each recording.}
%   \label{fig:raw-data-clusters}
% \end{figure}
Comparing our groups to those we would find by applying the same
procedure to
the distances between note-by-note tempo vectors reveals a number of
differences (see the Supplement~\citep{McDonaldMcBride2021a} for the distance matrix calculated in
this way). The four similar recordings in \autoref{fig:similar} would be
spread across three different groups, for example, as would our
group one. On the other hand, grouping by tempo vectors often (somewhat
miraculously) groups recordings by the same pianist together: both
recordings by Smidowicz (same as grouping by parameters), three of
the four recordings by Rubinstein, 
and both recordings by Hatto.  Both metrics see Cortot's recording as a
strong outlier (the remote \blkdiamond\ in \autoref{fig:pca}). In terms of Equation~\eqref{eq:parameter-distance}, Cortot's
recording is 1.7 times farther from it's nearest neighbor than is the
case for the
next most dissimilar recording.
\begin{figure}[t]
  \centering
  \includegraphics[width=.9\linewidth]{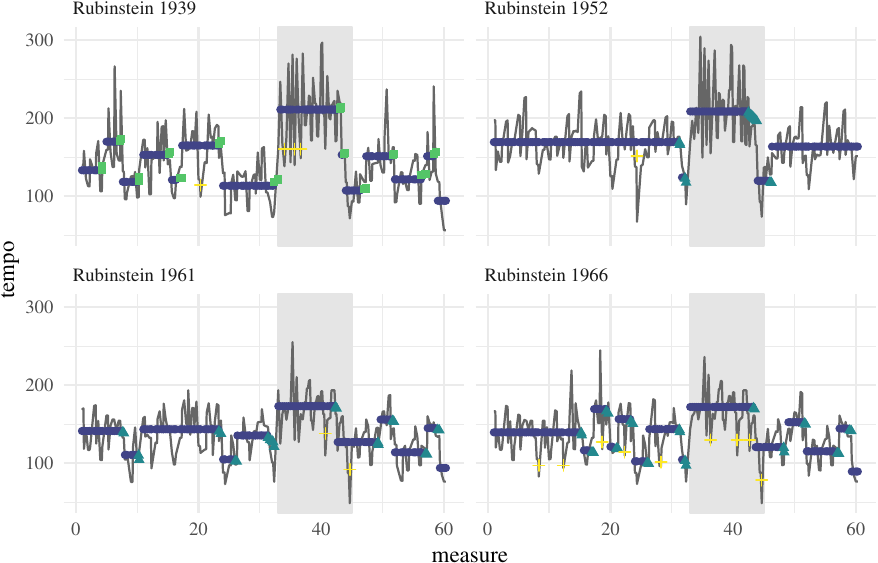}
  \caption{The four recordings by Arthur Rubinstein. Our clustering
    puts the 1952 and 1961 recordings in clusters one and two while
    leaving the others out. Clustering by tempo vector separates 1952
    from the other three.}
  \label{fig:rubinstein}
\end{figure}

\autoref{fig:rubinstein} shows all four Rubinstein recordings. The
1939 recording is rather odd in that the measures 24--32 are so slow
relative to the rest of the A section. The variability in the 1966
recording nearly obscures the contrast between the B section and the
surrounding A sections. These two recordings are nonetheless grouped
together by the tempo vectors. Our method on the other hand, puts
these four recordings in different groups. The estimated parameters for these four
performances are shown in the bottom half of
\autoref{tab:similar-different-perf}. The top half shows the
parameters for the four similar performances in
\autoref{fig:similar}. There is much larger variability across
Rubinstein's recordings, as we would expect.
\begin{table}[tb]
  \centering
  \caption{The estimated parameters for the four similar performances
    in group four and those for all four by Arthur Rubinstein.
    \label{tab:similar-different-perf}}
  \resizebox{\linewidth}{!}{%
  \begin{tabular}{@{}lrrrrrrrrrrrr@{}}
    \toprule
    & $\sigma^2_\epsilon$ & $\mu_{\textrm{tempo}}$
    & $\mu_{\textrm{acc}}$ & $\mu_{\textrm{stress}}$
    & $\sigma^2_{\textrm{tempo}}$ & $p_{11}$ & $p_{12}$ & $p_{22}$
    & $p_{31}$ & $p_{13}$ & $p_{21}$ & $p_{32}$\\
    \midrule
Wasowski 1990 & 414.99 & 132.00 & -10.00 & -40.00 & 425.00 & 0.85 & 0.05 & 0.67 & 0.34 & 0.02 & 0.26 & 0.2 \\    Shebanova 2002 & 439.98 & 132.00 & -10.00 & -40.00 & 400.02 & 0.85 & 0.05 & 0.67 & 0.33 & 0.02 & 0.27 & 0.20\\    Richter 1976 & 426.70 & 136.33 & -11.84 & -34.82 & 439.38 & 0.85 & 0.05 & 0.74 & 0.44 & 0.02 & 0.25 & 0.17\\    Milkina 1970 & 435.25 & 136.38 & -9.68 & -40.02 & 400.01 & 0.87 & 0.05 & 0.68 & 0.33 & 0.02 & 0.26 & 0.21\\    \midrule
Rubinstein 1939 & 520.32 & 145.26 & -7.89 & -50.82 & 345.64 & 0.89 & 0.02 & 0.83 & 0.56 & 0.05 & 0.13 & 0.16\\    Rubinstein 1952 & 481.13 & 128.13 & -7.76 & -17.59 & 409.30 & 0.93 & 0.04 & 0.68 & 0.32 & 0.01 & 0.28 & 0.19\\    Rubinstein 1961 & 434.23 & 139.17 & -8.34 & -35.08 & 355.00 & 0.90 & 0.06 & 0.56 & 0.46 & 0.01 & 0.41 & 0.19\\    Rubinstein 1966 & 380.95 & 127.24 & -8.80 & -42.28 & 473.69 & 0.87 & 0.07 & 0.36 & 0.34 & 0.01 & 0.61 & 0.20\\
    \bottomrule
  \end{tabular}
  }
\end{table}

\subsection{Alternative smoothers}
\label{sec:altern-smooth}

Our model is just one type of smoothing one could imagine using to
find low-dimensional structure for the vector of note-by-note
tempos. Alternative statistical techniques are common, and examining
how they compare with our method helps to illuminate some of its
benefits. The most obvious alternative is to use 
splines~\citep{CravenWahba1978,Wahba1990} though total-variation
denoising or trend filtering \citep{KimKoh2009,Tibshirani2014} are
other reasonable alternatives.
These statistical techniques perform smoothing by encouraging small
changes in derivatives (splines) or bounded total variation
(trend filtering). 
But musical performances do not conform to these assumptions because tempo interpretations rely on the juxtaposition of local smoothness
with sudden changes and emphases to create listener interest. It is
exactly the parts of a performance that are poorly described by
statistical smoothers that render a performance
interesting. Furthermore, many of these
inflections are notated by the 
composer or are implicit in performance practice developed over
centuries of musical expressivity. Consequently, smoothing that
incorporates domain knowledge leads to better statistical and
empirical results.
\begin{figure}[t]
  \centering
  \includegraphics[width=.9\linewidth]{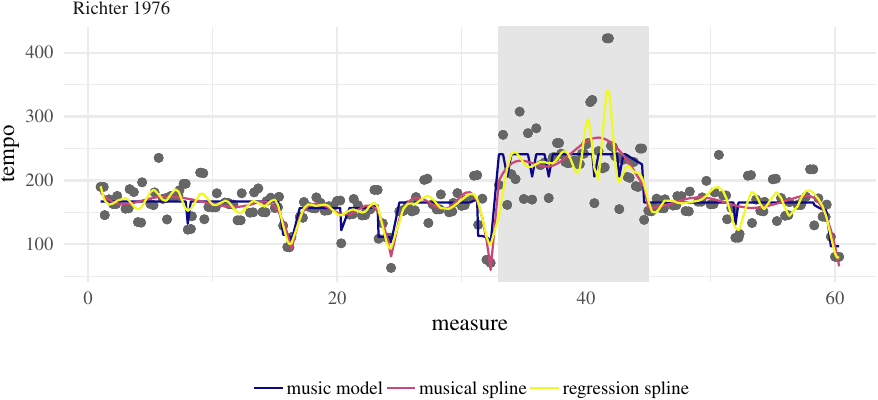}
  \caption{Smoothing with splines and musical models}
  \label{fig:splines}
\end{figure}

\autoref{fig:splines}
shows the note-by-note tempo of Richter's 1976 recording. Regression
splines with 
equally spaced knots are shown with a dashed line. We use generalized cross
validation~\citep{GolubHeath1979} to select the number of knots (one
knot per measure). The dotted line shows a regression spline fewer knots,
but whose locations were chosen
manually
to coincide with the musical phrase endings discussed in
\autoref{sec:musical-analysis}. Knots at phrase endings were
duplicated up to four times to allow for discontinuities. The solid line 
shows the estimated smooth tempo from our model (the same as in
\autoref{fig:archetypal}). The regression spline with
equally spaced knots undersmooths in constant tempo areas in an
attempt to capture sudden emphases and dramatic changes in others. The spline
with informed knot choice does much better, picking up the periods of
deceleration at the ends of phrases. Our model learns these behaviors
on its own while also capturing individual emphases that are missed in
the musical analysis but are idiosyncratic to Richter's playing. It is
also more parsimonious to musical interpretation, inferring constant
tempo periods rather than resulting in smoothly varying tempos in
stable periods.%, such as measures 1--16.

\subsection{Problems with the model and estimation}
\label{sec:problems-with-model}

While our model of musical decision making yields interesting insights
into performance practice most of the time, it also suffers from some
deficiencies. As discussed above in reference to Alfred Cortot's
recording, the assumption that all parameters are stable over the
entire piece may not always be accurate. The $\mu_{\textrm{tempo}}$
parameter especially, should be estimated separately in different
sections. This problem will only be compounded in more complex music
with many contrasting sections. A related issue is the current form
for the slowing down and speeding up sections. Our model assumes that
both occur linearly, with a constant decrease of $\mu_{\textrm{acc}}$
b.p.m. An ability to slow increasingly as one remains in the state may
improve the model fit. The multiplicative model described in more
detail in the Supplementary Material addresses this issue but
requires further work.

There is nothing intrinsic to the model which forces states 2,
3, or 4 to always go in the correct direction. If for example,
$\mu_{\textrm{acc}}$ is small in magnitude relative to
$\sigma^2_{\textrm{acc}}$, a purposeful acceleration could be explained as time
spent in state 2 but with large positive errors. For this piece, the
priors help to avoid such occurrences, but this aspect of the
Gaussian state-space model could be improved by enforcing non-Gaussian
behavior. Of course, such constraints would complicate likelihood
evaluation since the Kalman filter could no longer be used. 

Relatedly,
our model produced objectively incorrect inferences on two performances
\begin{figure}[t]
  \centering
  \includegraphics[width=.9\linewidth]{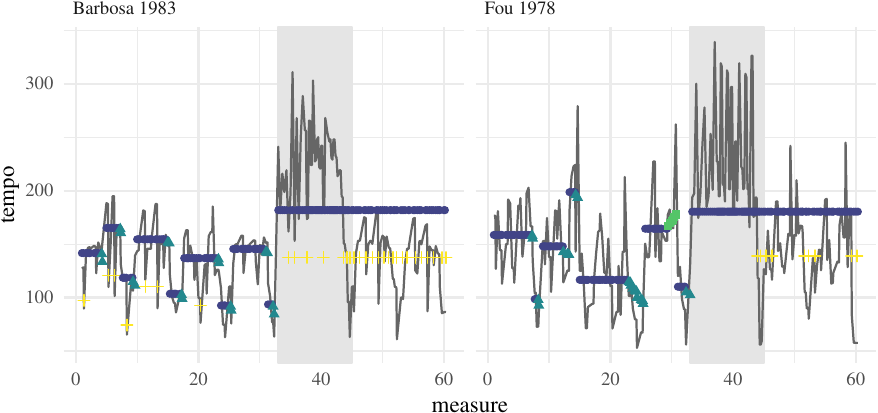}
  \caption{Estimation errors on two performances.}
  \label{fig:bad-model}
\end{figure}
(\autoref{fig:bad-model}). Here, the estimated
path failed to transition to state 1 at the
recapitulation of the A section. In both cases, the resulting path
stands out dramatically, remaining in the much faster constant tempo
state from the B section with overly frequent emphases. Both of these
performances are quite volatile, making estimation difficult. Altering
the prior distributions along the lines suggested in the next section
may help.

\subsection{Prior sensitivity and generalization}
\label{sec:prior-sens-gener}

As discussed in \autoref{sec:penal-maxim-likel}, the main reason for the prior distributions shown in
\autoref{tab:priors} is that they help to identify the
parameters. It is this identifiability issue that mainly guided our
choices. For instance, if $\mu_{\textrm{stress}}$ is too similar to
$\mu_{\textrm{acc}}$, then a sequence like $1\rightarrow 4 \rightarrow
1 \rightarrow 1$ will be hard to distinguish from $1\rightarrow 2
\rightarrow 2 \rightarrow 1$. So, while it is important that those
priors have low probability on common support, their shapes are less important for
inference. Similar arguments hold for the other parameters.

With this intuition in mind, these priors should allow the model to
work reasonably on similar piano pieces, even from different eras
(Baroque, Classical, Modernist). That said, the specific
values of prior means (for example, 10 for $\mu_{\mbox{acc}}$)
would be better expressed relative to the overall average speed,
perhaps as $\overline{Y}/10$ b.p.m.\ or similar. In this way, really
slow or fast pieces could be more easily accommodated. Since we only looked at
one 
score, it is sufficient to simply fix some values. Using these 4
states should be enough for many types of music, even beyond piano
recordings. However, music with more
sections, say a piano sonata or a Prelude by Claude Debussy, would
likely benefit from the inclusion of more discrete states. Adding
another layer to \autoref{fig:switchss} that can handle formal
divisions is one such option, while employing a Dirichlet process
similar to that used by \citet{RenDunson2010} may also work.

To gauge prior sensitivity (subject to the constraints above), we also
estimated the model under alternative specifications. In particular we
examine Sviatoslav Richter's 
1976 performance under four alternative priors: (1) replacing
all Gamma distributions with inverse Gamma to examine the influence
of tail shape; (2) making $\sigma^2_\epsilon$ smaller; (3) setting
$p(\sigma^2_\epsilon)=p(\sigma^2_{\textrm{tempo}}) \propto 1$; (4)
replacing the informative transition probability priors with uniform
distributions. We calculate the root-mean squared error (RMSE) and
the negative log likelihood under different choices. The results are
roughly similar both in terms of quantitative metrics and the
qualitative inferences based on the figures. The most different
setting occurs for choice (2) which decreases the quality of the fit,
eliminates the use of the ``emphasis'' state, and has trouble dealing
with the faster B section. A more comprehensive evaluation of these prior
choices is included in the 
Supplement where we show the specific distributions and the
inferred performance decisions (like in \autoref{fig:archetypal})
under each choice.

\section{Discussion}
\label{sec:discussion}

Musical interpretation is the most important factor 
in determining whether or not concertgoers enjoy a classical performance. Every
performance includes mistakes---intonation issues, a lost note, an
unpleasant sound---but these are all easily forgotten (or unnoticed) when a performer
engages her audience, imbuing a piece with novel emotional content
beyond the vague instructions inscribed on the printed page. While music teachers use
imagery or heuristic guidelines to motivate interpretive decisions, combining these
vague instructions to create a convincing performance remains the domain
of the performer, subject to the whims of the moment, technical
fluency, and taste.

In this paper, we develop a statistical model for tempo to elucidate performance
decisions from classical music recordings. We present an algorithm for
performing likelihood inference, estimate our model using a large
collection of recordings of the same composition, and demonstrate how
the model is able to recover performer intentions, and how they relate
to standard musical analysis. While our methods perform well, our
analysis reveals a number of avenues for future work and
improvement. For the piano, apart from tempo decisions, the performer
can also control dynamics differentially. Similar techniques to those
employed here could be used to describe levels of loudness, and
creating a model that combined both is desirable. Pianists have
relatively few variables under their control for interpretation:
tempo, dynamics, and pedalling. On the other hand, string players have
many more. Bowing decisions, fingerings, vibrato, and broken chords are
all important tools which are difficult to discern aurally from a recording, let
alone describe with a simple statistical model. Significant work would
be required to generalize our techniques to more detailed
interpretative analysis. Examining more complex genres---sonatas, string quartets,
symphonies---would also be interesting for future work.

Another avenue we wish to pursue in the future is to examine how our
model's implications may be useful for teaching students. Can we
estimate it quickly to provide immediate feedback to novice pianists?
In this paper, we used a dataset in which the note-by-note tempos were
annotated by experienced musicians. Combining our model with existing
approaches to solving the note-score alignment problem
\citep{LangFreitas2005,Raphael2002,DannenbergRaphael2006}, perhaps to
their benefit would be the first step. Together, this could produce an
immediate graphical representation that students and teachers could
use to evaluate and improve their practice.

\section*{Acknowledgements}

DJM was partially supported by the National Science Foundation Grant Nos.\
DMS--1407439 and DMS--1753171. CR was partially supported by National Science
Foundation Grants IIS--1526473 and  IIS--0812244.

\clearpage

\bibliographystyle{rss}
\bibliography{chopinrefs}

\begin{thebibliography}{74}
\expandafter\ifx\csname natexlab\endcsname\relax\def\natexlab#1{#1}\fi
\expandafter\ifx\csname url\endcsname\relax
  \def\url#1{\texttt{#1}}\fi
\expandafter\ifx\csname urlprefix\endcsname\relax\def\urlprefix{URL: }\fi

\bibitem[{Anderson and Moore(1979)}]{AndersonMoore1979}
Anderson, B.~D. and Moore, J.~B. (1979) \textit{Optimal filtering}.
\newblock Englewood Cliffs, NJ: Prentice-Hall.

\bibitem[{Andrieu et~al.(2010)Andrieu, Doucet and Holenstein}]{AndrieuDoucet2010}
Andrieu, C., Doucet, A. and Holenstein, R. (2010) Particle {M}arkov chain {Monte C}arlo methods.
\newblock \textit{Journal of the Royal Statistical Society. Series B, Statistical Methodology}, \textbf{72}, 1--33.

\bibitem[{Arcos and Mantaras(2001)}]{Arcos}
Arcos, J. and Mantaras, R.~L. (2001) An interactive cbr approach for generating expressive music.
\newblock \textit{Journal of Applied Intelligence}, \textbf{21}, 115--129.

\bibitem[{Ariza(2005)}]{Ariza2005}
Ariza, C. (2005) Navigating the landscape of computer aided algorithmic composition systems: a definition, seven descriptors, and a lexicon of systems and research.
\newblock In \textit{Proceedings of International Computer Music Conference}.

\bibitem[{Arzt and Widmer(2015)}]{ArztWidmer2015}
Arzt, A. and Widmer, G. (2015) Real-time music tracking using multiple performances as a reference.
\newblock In \textit{International Society for Music Information Retrieval (ISMIR)}, 357--363.

\bibitem[{Bernstein(2005)}]{Bernstein2005}
Bernstein, L. (2005) \textit{Young People's Concerts}.
\newblock Pompton Plains, NJ: Amadeus Press.

\bibitem[{Bisiani(1992)}]{Bisiani1992}
Bisiani, R. (1992) Beam search.
\newblock In \textit{Encyclopedia of Artificial Intelligence} (ed. S.~Shapiro). John Wiley and Sons, 2nd edn.

\bibitem[{Block et~al.(2011)Block, Jonsen, Jorgensen, Winship, Shaffer, Bograd, Hazen, Foley, Breed, Harrison et~al.}]{BlockJonsen2011}
Block, B.~A., Jonsen, I.~D., Jorgensen, S.~J., Winship, A.~J., Shaffer, S.~A., Bograd, S.~J., Hazen, E.~L., Foley, D.~G., Breed, G., Harrison, A.-L. et~al. (2011) Tracking apex marine predator movements in a dynamic ocean.
\newblock \textit{Nature}, \textbf{475}, 86.

\bibitem[{Boulanger-Lewandowski et~al.(2012)Boulanger-Lewandowski, Bengio and Vincent}]{Boulanger-LewandowskiBengio2012}
Boulanger-Lewandowski, N., Bengio, Y. and Vincent, P. (2012) Modeling temporal dependencies in high-dimensional sequences: Application to polyphonic music generation and transcription.
\newblock In \textit{Proceedings of the 29th International Conference on Machine Learning}. Edinburgh, Scotland, UK.

\bibitem[{Bresin et~al.(2002)Bresin, Friberg and Sundberg}]{Bresin}
Bresin, R., Friberg, A. and Sundberg, J. (2002) Director musices: The {KTH} performance rules system.
\newblock In \textit{Proceedings of SIGMUS-46}. Kyoto.

\bibitem[{Burkholder et~al.(2014)Burkholder, Grout and Palisca}]{BurkholderGrout2014}
Burkholder, J.~P., Grout, D.~J. and Palisca, C.~V. (2014) \textit{A History of Western Music}.
\newblock WW Norton \& Company, 9th edn.

\bibitem[{CHARM(2009)}]{CHARM-site}
CHARM (2009) {Centre for the History and Analysis of Recorded Music}.
\newblock \urlprefix\url{http://www.charm.rhul.ac.uk/about/about.html}.
\newblock Online; accessed 12 March 2019.

\bibitem[{Collins(2016)}]{Collins2016}
Collins, N. (2016) A funny thing happened on the way to the formula: Algorithmic composition for musical theater.
\newblock \textit{Computer Music Journal}, \textbf{40}, 41--57.

\bibitem[{Cont(2010)}]{Cont2010}
Cont, A. (2010) A coupled duration-focused architecture for real-time music-to-score alignment.
\newblock \textit{IEEE Transactions on Pattern Analysis and Machine Intelligence}, \textbf{32}, 974--987.

\bibitem[{Cont et~al.(2007)Cont, Schwarz, Schnell and Raphael}]{ContSchwarz2007}
Cont, A., Schwarz, D., Schnell, N. and Raphael, C. (2007) Evaluation of real-time audio-to-score alignment.
\newblock In \textit{International Symposium on Music Information Retrieval (ISMIR)}.

\bibitem[{Cook(2013)}]{Cook2013}
Cook, N. (2013) \textit{Beyond the score: Music as performance}.
\newblock Oxford University Press.

\bibitem[{Craven and Wahba(1978)}]{CravenWahba1978}
Craven, P. and Wahba, G. (1978) Smoothing noisy data with spline functions.
\newblock \textit{Numerische Mathematik}, \textbf{31}, 377--403.

\bibitem[{Dannenberg(1985)}]{Dannenberg1985}
Dannenberg, R. (1985) An on-line algorithm for real-time accompaniment.
\newblock In \textit{Proceedings of the 1984 International Computer Music Conference}, 193--198. International Computer Music Association.

\bibitem[{Dannenberg and Raphael(2006)}]{DannenbergRaphael2006}
Dannenberg, R.~B. and Raphael, C. (2006) Music score alignment and computer accompaniment.
\newblock \textit{Communications of the ACM}, \textbf{49}, 38--43.

\bibitem[{Dror et~al.(2012)Dror, Koenigstein, Koren and Weimer}]{DrorKoenigstein2012}
Dror, G., Koenigstein, N., Koren, Y. and Weimer, M. (2012) The {Yahoo! Music Dataset} and {KDD-Cup'11}.
\newblock In \textit{KDD Cup}, 8--18.

\bibitem[{Dudoit and Fridlyand(2002)}]{DudoitFridlyand2002}
Dudoit, S. and Fridlyand, J. (2002) A prediction-based resampling method for estimating the number of clusters in a dataset.
\newblock \textit{Genome Biology}, \textbf{3}, research0036.1.

\bibitem[{Durbin and Koopman(2001)}]{DurbinKoopman2001}
Durbin, J. and Koopman, S. (2001) \textit{Time Series Analysis by State Space Methods}.
\newblock Oxford: Oxford Univ Press.

\bibitem[{Durbin and Koopman(1997)}]{DurbinKoopman1997}
Durbin, J. and Koopman, S.~J. (1997) {Monte Carlo} maximum likelihood extimation for non-{G}aussian state space models.
\newblock \textit{Biometrika}, \textbf{84}, 669--684.

\bibitem[{Earis(2007)}]{Earis2007}
Earis, A. (2007) An algorithm to extract expressive timing and dynamics from piano recordings.
\newblock \textit{Musicae Scientiae}, \textbf{11}, 155--182.

\bibitem[{Earis(2009)}]{Earis2009}
--- (2009) {Mazurka in F Major, Op. 68, No. 3}.
\newblock \urlprefix\url{http://mazurka.org.uk/auto/earis/mazurka68-3/}.
\newblock Accessed 12 March 2019.

\bibitem[{Eddelbuettel(2013)}]{Eddelbuettel2013}
Eddelbuettel, D. (2013) \textit{Seamless {R} and {C++} Integration with {Rcpp}}.
\newblock New York: Springer.

\bibitem[{Fearnhead and Clifford(2003)}]{FearnheadClifford2003}
Fearnhead, P. and Clifford, P. (2003) On-line inference for hidden markov models via particle filters.
\newblock \textit{Journal of the Royal Statistical Society: Series B (Statistical Methodology)}, \textbf{65}, 887--899.

\bibitem[{Flossman S.~Grachten and Widmer(2012)}]{Flossman}
Flossman S.~Grachten, M. and Widmer, G. (2012) Expressive performance rendering with probabilistic models.
\newblock In \textit{Guide to Computing for Expressive Music Performance} (eds. A.~Kirke and E.~Miranda). Springer.

\bibitem[{Flossmann et~al.(2013)Flossmann, Grachten and Widmer}]{FlossmannGrachten2013}
Flossmann, S., Grachten, M. and Widmer, G. (2013) Expressive performance rendering with probabilistic models.
\newblock In \textit{Guide to Computing for Expressive Music Performance}, 75--98. Springer.

\bibitem[{Fors{\'e}n et~al.(2013)Fors{\'e}n, Gray, Lindgren and Gray}]{ForsenGray2013}
Fors{\'e}n, S., Gray, H.~B., Lindgren, L.~O. and Gray, S.~B. (2013) Was something wrong with {B}eethoven's metronome?
\newblock \textit{Notices of the AMS}, \textbf{60}.

\bibitem[{Fox et~al.(2011)Fox, Sudderth, Jordan and Willsky}]{FoxSudderth2011}
Fox, E.~B., Sudderth, E.~B., Jordan, M.~I. and Willsky, A.~S. (2011) A sticky {HDP-HMM} with application to speaker diarization.
\newblock \textit{The Annals of Applied Statistics}, \textbf{5}, 1020--1056.

\bibitem[{Fuh(2006)}]{Fuh2006}
Fuh, C.-D. (2006) Efficient likelihood estimation in state space models.
\newblock \textit{Annals of Statistics}, \textbf{34}, 2026--2068.

\bibitem[{Ghahramani and Hinton(2000)}]{GhahramaniHinton2000}
Ghahramani, Z. and Hinton, G.~E. (2000) Variational learning for switching state-space models.
\newblock \textit{Neural Computation}, \textbf{12}, 831--864.

\bibitem[{Golub et~al.(1979)Golub, Heath and Wahba}]{GolubHeath1979}
Golub, G.~H., Heath, M. and Wahba, G. (1979) Generalized cross-validation as a method for choosing a good ridge parameter.
\newblock \textit{Technometrics}, \textbf{21}, 215--223.

\bibitem[{Grindlay and Helmbold(2006)}]{Grindlay}
Grindlay, G. and Helmbold, D. (2006) Modeling, analyzing, and synthesizing expressive piano performance with graphical models.
\newblock \textit{Machine Learning}, \textbf{65}, 361--387.

\bibitem[{Gu and Raphael(2012)}]{GuRaphael2012}
Gu, Y. and Raphael, C. (2012) Modeling piano interpretation using switching {K}alman filter.
\newblock In \textit{International Society for Music Information Retrieval (ISMIR)}, 145--150.

\bibitem[{Hadjeres et~al.(2017)Hadjeres, Pachet and Nielsen}]{HadjeresPachet2017}
Hadjeres, G., Pachet, F. and Nielsen, F. (2017) {D}eep{B}ach: a steerable model for {B}ach chorales generation.
\newblock In \textit{Proceedings of the 34th International Conference on Machine Learning} (eds. D.~Precup and Y.~W. Teh), vol.~70 of \textit{Proceedings of Machine Learning Research}, 1362--1371. International Convention Centre, Sydney, Australia: PMLR.

\bibitem[{Hamilton(2011)}]{Hamilton2011}
Hamilton, J. (2011) Calling recessions in real time.
\newblock \textit{International Journal of Forecasting}, \textbf{27}, 1006--126.

\bibitem[{Harvey(1990)}]{Harvey1990}
Harvey, A.~C. (1990) \textit{Forecasting, structural time series models and the Kalman filter}.
\newblock Cambridge University Press.

\bibitem[{Kallberg(1996)}]{Kallberg1996}
Kallberg, J. (1996) \textit{Chopin at the boundaries: Sex, history, and musical genre}.
\newblock Harvard University Press.

\bibitem[{Kalman(1960)}]{Kalman1960}
Kalman, R.~E. (1960) A new approach to linear filtering and prediction problems.
\newblock \textit{Journal of Basic Engineering}, \textbf{82}, 35--45.

\bibitem[{Kim and Nelson(1998)}]{KimNelson1998}
Kim, C. and Nelson, C. (1998) Business cycle turning points, a new coincident index, and tests of duration dependence based on a dynamic factor model with regime switching.
\newblock \textit{Review of Economics and Statistics}, \textbf{80}, 188--201.

\bibitem[{Kim(1994)}]{Kim1994}
Kim, C.-J. (1994) Dynamic linear models with markov-switching.
\newblock \textit{Journal of Econometrics}, \textbf{60}, 1--22.

\bibitem[{Kim et~al.(2009)Kim, Koh, Boyd and Gorinevsky}]{KimKoh2009}
Kim, S.-J., Koh, K., Boyd, S. and Gorinevsky, D. (2009) $\ell_1$ trend filtering.
\newblock \textit{SIAM Review}, \textbf{51}, 339--360.

\bibitem[{Kitagawa(1987)}]{Kitagawa1987}
Kitagawa, G. (1987) Non-{G}aussian state-space modeling of nonstationary time series.
\newblock \textit{Journal of the American Statistical Association}, \textbf{82}, 1032--1041.

\bibitem[{Kitagawa(1996)}]{Kitagawa1996}
--- (1996) {Monte C}arlo filter and smoother for non-{G}aussian nonlinear state space models.
\newblock \textit{Journal of Computational and Graphical Statistics}, 1--25.

\bibitem[{Koyama et~al.(2010)Koyama, P{\'e}rez-Bolde, Shalizi and Kass}]{KoyamaPerez-Bolde2010}
Koyama, S., P{\'e}rez-Bolde, L.~C., Shalizi, C.~R. and Kass, R.~E. (2010) Approximate methods for state-space models.
\newblock \textit{Journal of the American Statistical Association}, \textbf{105}, 170--180.

\bibitem[{Lang and Freitas(2005)}]{LangFreitas2005}
Lang, D. and Freitas, N.~D. (2005) Beat tracking the graphical model way.
\newblock In \textit{Advances in Neural Information Processing Systems}, 745--752. Cambridge, MA: MIT press.

\bibitem[{Lang et~al.(2017)Lang, Bischl and Surmann}]{LangBischl2017}
Lang, M., Bischl, B. and Surmann, D. (2017) batchtools: Tools for {R} to work on batch systems.
\newblock \textit{The Journal of Open Source Software}, \textbf{2}, 135.

\bibitem[{Maezawa(2019)}]{Maezawa}
Maezawa, A. (2019) Deep linear autoregressive model of interpretable prediction of expressive tempo.
\newblock In \textit{Proceedings of the 16th Sound and Music Computing Conference}. M{\'a}laga, Spain.

\bibitem[{McDonald et~al.(2021)McDonald, McBride, Gu and Raphael}]{McDonaldMcBride2021a}
McDonald, D.~J., McBride, M., Gu, Y. and Raphael, C. (2021) Supplement to "{M}arkov-switching state space models for uncovering musical interpretation".
\newblock URL: \url{https://doi.org/10.1214/21-AOAS1457SUPPA}, \url{https://doi.org/10.1214/21-AOAS1457SUPPB}, \url{https://doi.org/10.1214/21-AOAS1457SUPPC}.

\bibitem[{McFee and Lanckriet(2011)}]{McFeeLanckriet2011}
McFee, B. and Lanckriet, G. (2011) Learning multi-modal similarity.
\newblock \textit{Journal of Machine Learning Research}, \textbf{12}, 491--523.

\bibitem[{Mead(2007)}]{Mead2007}
Mead, A. (2007) On tempo relations.
\newblock \textit{Perspectives of New Music}, \textbf{45}, 64--108.

\bibitem[{van~den Oord et~al.(2013)van~den Oord, Dieleman and Schrauwen}]{OordDieleman2013}
van~den Oord, A., Dieleman, S. and Schrauwen, B. (2013) Deep content-based music recommendation.
\newblock In \textit{Advances in Neural Information Processing Systems 26} (eds. C.~J.~C. Burges, L.~Bottou, M.~Welling, Z.~Ghahramani and K.~Q. Weinberger), 2643--2651. Curran Associates, Inc.

\bibitem[{Patterson et~al.(2008)Patterson, Thomas, Wilcox, Ovaskainen and Matthiopoulos}]{PattersonThomas2008}
Patterson, T.~A., Thomas, L., Wilcox, C., Ovaskainen, O. and Matthiopoulos, J. (2008) State--space models of individual animal movement.
\newblock \textit{Trends in ecology \& evolution}, \textbf{23}, 87--94.

\bibitem[{{R Core Team}(2019)}]{R-Core-Team2019}
{R Core Team} (2019) \textit{R: A Language and Environment for Statistical Computing}.
\newblock R Foundation for Statistical Computing, Vienna, Austria.
\newblock \urlprefix\url{https://www.R-project.org/}.

\bibitem[{Raphael(2002)}]{Raphael2002}
Raphael, C. (2002) A hybrid graphical model for rhythmic parsing.
\newblock \textit{Artificial Intelligence}, \textbf{137}, 217--238.

\bibitem[{Raphael(2010)}]{Raphael2010}
--- (2010) Music plus one and machine learning.
\newblock In \textit{Proceedings of the 27th International Conference on Machine Learning (ICML-10)} (eds. J.~F{\"u}rnkranz and T.~Joachims), 21--28. Haifa, Israel.

\bibitem[{Rauch et~al.(1965)Rauch, Striebel and Tung}]{RauchStriebel1965}
Rauch, H.~E., Striebel, C. and Tung, F. (1965) Maximum likelihood estimates of linear dynamic systems.
\newblock \textit{AIAA journal}, \textbf{3}, 1445--1450.

\bibitem[{Ren et~al.(2010)Ren, Dunson, Lindroth and Carin}]{RenDunson2010}
Ren, L., Dunson, D., Lindroth, S. and Carin, L. (2010) Dynamic nonparametric {B}ayesian models for analysis of music.
\newblock \textit{Journal of the American Statistical Association}, \textbf{105}, 458--472.

\bibitem[{Roberts et~al.(2018{\natexlab{a}})Roberts, Engel, Raffel, Hawthorne and Eck}]{RobertsEngel2018}
Roberts, A., Engel, J., Raffel, C., Hawthorne, C. and Eck, D. (2018{\natexlab{a}}) A hierarchical latent vector model for learning long-term structure in music.
\newblock In \textit{Proceedings of the 35th International Conference on Machine Learning} (eds. J.~Dy and A.~Krause), vol.~80 of \textit{Proceedings of Machine Learning Research}, 4364--4373. Stockholmsm{\"a}ssan, Stockholm Sweden: PMLR.

\bibitem[{Roberts et~al.(2018{\natexlab{b}})Roberts, Hawthorne and Simon}]{RobertsHawthorne2018}
Roberts, A., Hawthorne, C. and Simon, I. (2018{\natexlab{b}}) Magenta.js: A javascript api for augmenting creativity with deep learning.
\newblock In \textit{Joint Workshop on Machine Learning for Music (ICML)}.

\bibitem[{Schedl et~al.(2014)Schedl, G{\'o}mez, Urbano et~al.}]{schedl2014music}
Schedl, M., G{\'o}mez, E., Urbano, J. et~al. (2014) Music information retrieval: Recent developments and applications.
\newblock \textit{Foundations and Trends{\textregistered} in Information Retrieval}, \textbf{8}, 127--261.

\bibitem[{Stowell and Chew(2012)}]{Chew}
Stowell, D. and Chew, E. (2012) {B}ayesian {MAP} estimation of piecewise arcs in tempo time series.
\newblock In \textit{Proceedings of Computer Music Multidisciplinary Research}.

\bibitem[{Sturm et~al.(2019)Sturm, Ben-Tal, Monaghan, Collins, Herremans, Chew, Hadjeres, Deruty and Pachet}]{SturmBen-Tal2019}
Sturm, B.~L., Ben-Tal, O., Monaghan, {\'U}., Collins, N., Herremans, D., Chew, E., Hadjeres, G., Deruty, E. and Pachet, F. (2019) Machine learning research that matters for music creation: A case study.
\newblock \textit{Journal of New Music Research}, \textbf{48}, 36--55.

\bibitem[{Thickstun et~al.(2017)Thickstun, Harchaoui and Kakade}]{ThickstunHarchaoui2017}
Thickstun, J., Harchaoui, Z. and Kakade, S.~M. (2017) Learning features of music from scratch.
\newblock In \textit{International Conference on Learning Representations (ICLR)}.

\bibitem[{Tibshirani et~al.(2001)Tibshirani, Walther and Hastie}]{TibshiraniWalther2001}
Tibshirani, R., Walther, G. and Hastie, T. (2001) Estimating the number of clusters in a data set via the gap statistic.
\newblock \textit{Journal of the Royal Statistical Society. Series B (Statistical Methodology)}, \textbf{63}, 411--423.

\bibitem[{Tibshirani(2014)}]{Tibshirani2014}
Tibshirani, R.~J. (2014) Adaptive piecewise polynomial estimation via trend filtering.
\newblock \textit{Annals of Statistics}, \textbf{42}, 285--323.

\bibitem[{Vercoe(1984)}]{Vercoe1985}
Vercoe, B. (1984) The synthetic performer in the context of live performance.
\newblock In \textit{Proceedings of the 1984 International Computer Music Conference}, 199--200. International Computer Music Association.

\bibitem[{Wahba(1990)}]{Wahba1990}
Wahba, G. (1990) \textit{Spline models for observational data}, vol.~59 of \textit{CBMS-NSF Regional Conference Series in Applied Mathematics}.
\newblock Philadelphia, PA: Society for Industrial and Applied Mathematics (SIAM).

\bibitem[{Whiteley et~al.(2010)Whiteley, Andrieu and Doucet}]{WhiteleyAndrieu2010}
Whiteley, N., Andrieu, C. and Doucet, A. (2010) Efficient {B}ayesian inference for switching state-space models using discrete particle {M}arkov chain {M}onte {C}arlo methods.
\newblock \textit{Tech. Rep. 10:04}, Bristol University.

\bibitem[{Wickham(2016)}]{Wickham2016}
Wickham, H. (2016) \textit{ggplot2: Elegant graphics for data analysis}.
\newblock Springer, 2nd edn.

\bibitem[{Wickham(2017)}]{Wickham2017}
--- (2017) \textit{tidyverse: Easily Install and Load the `Tidyverse'}.
\newblock \urlprefix\url{https://CRAN.R-project.org/package=tidyverse}.
\newblock R package version 1.2.1.

\bibitem[{Widmer et~al.(2009)Widmer, Flossmann and Grachten}]{Widmer}
Widmer, G., Flossmann, S. and Grachten, M. (2009) {YQX} plays chopin.
\newblock \textit{AI Magazine}, \textbf{30}, 35.

\end{thebibliography}

\clearpage

\appendix

\section{Supplementary material}

\hypertarget{algorithms}{%
\subsection{Algorithms}\label{algorithms}}

For completeness, we include here concise descriptions of the Kalman
filter and smoother we employ as inputs to our main algorithm. The
filter is given in \autoref{alg:kalman}.

\begin{algorithm}
  \caption{Kalman filter: estimate $x_i$ conditional on
    $\{y_j\}_{j=1}^i$, for all $i=1,\ldots,n$ and calculate the log likelihood
    for $\theta$\label{alg:kalman}}
  \begin{algorithmic}
    \STATE {\bf Input:} $Y$, $x_0$, $P_0$, $d,\ T,\ c,\ Z,$ and $G$
    \STATE $\ell(\theta) \leftarrow 0$ \COMMENT{Initialize the log-likelihood}
    \FOR{$i=1$ to  $n$}
    \STATE $\begin{aligned}\varx_{i}
      &\leftarrow d + T x_{i-1|i-1}, & P_i &\leftarrow Q + T P_{i-1|i-1}
      T^\top\end{aligned}$ \COMMENT{Predict current state}
    \STATE $\begin{aligned}\widetilde{y}_i
      &\leftarrow c + Z \varx_i, & F_i &\leftarrow G + Z P_i
      Z^\top\end{aligned}$ \COMMENT{Predict current observation}
    \STATE $\begin{aligned}v_i&\leftarrow y_i-\widetilde{y}_i& K_i&
      \leftarrow P_i Z^\top F^{-1}\end{aligned}$ \COMMENT{Forecast error and 
    Kalman gain}
    \STATE $\begin{aligned} x_{i|i}
      &\leftarrow \varx_i + K_i v_i, & P_{i|i} &\leftarrow P_i - P_iZ^\top
      K_i\end{aligned}$ \COMMENT{Update}
    \STATE $\ell(\theta) = \ell(\theta) -v_i^\top F^{-1}v_i - \log(|F_i|)$
    \ENDFOR
    \RETURN $\widetilde{Y}=\{\widetilde{y}_i\}_{i=1}^n,\ \varx=\{\varx_i\}_{i=1}^n,\
    \widetilde{X}=\{x_{i|i}\}_{i=1}^n,\ P=\{P_i\}_{i=1}^n,\
    \widetilde{P}=\{P_{i|i}\}_{i=1}^n,\ \ell(\theta)$
  \end{algorithmic}
\end{algorithm}

To incorporate all future observations into these estimates, the Kalman
smoother is required. There are many different smoother algorithms
tailored for different applications. \autoref{alg:kalman-smoother}, due
to \citet{RauchStriebel1965}, is often referred to as the classical
fixed-interval smoother \citep{AndersonMoore1979}. It produces only the
unconditional expectations of the hidden state
\(\hat{x}_i=\Expect{x_i\given y_1,\ldots,y_n}\) for the sake of
computational speed. This version is more appropriate for inference in
the type of switching models we discuss in the manuscript.

\begin{algorithm}
  \caption{Kalman smoother (Rauch-Tung-Striebel): estimate $\hat{X}$ conditional on
    $Y$\label{alg:kalman-smoother}} 
  \begin{algorithmic}
    \STATE {\bf Input:} $\varx$, $\widetilde{X}$, $P$, $\widetilde{P}$,
    $T,$ $c$, $Z$.
    \STATE $i=n$,
    \STATE $\hat{x}_{n}\leftarrow \widetilde{x}_n$, 
    \WHILE{$t>1$}
    \STATE $\hat{y}_i \leftarrow c + Z\hat{x}_i,$
    \COMMENT{Predict observation vector}
    \STATE $\begin{aligned} e &\leftarrow \hat{x}_i -
      \varx_i, & V &\leftarrow P_i^{-1}\end{aligned}$,
    \STATE $i\leftarrow i-1$, \COMMENT{Increment}
    \STATE $\hat{x}_i = \widetilde{x}_i + \widetilde{P}_i T Ve $ 
    \ENDWHILE
    \RETURN $\widehat{Y}=\{\hat{y}_i\}_{i=1}^n, \hat{X}=\{\hat{x}_i\}_{i=1}^n$
  \end{algorithmic}
\end{algorithm}

\hypertarget{principal-components}{%
\subsection{Principal components}\label{principal-components}}

\begin{figure}

{\centering \includegraphics[width=5in,height=5in]{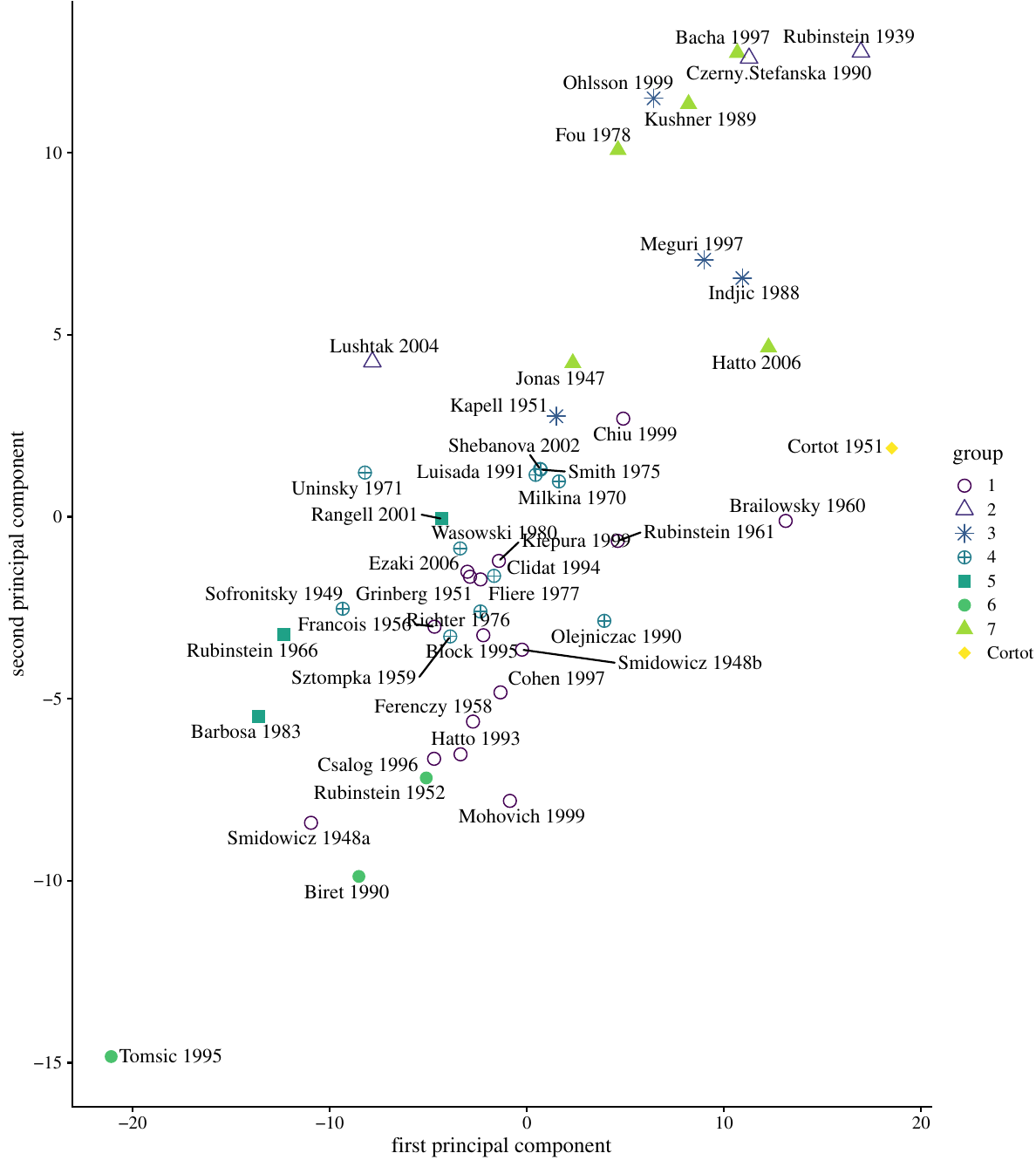} 

}

\caption{The first two principal components of the matrix of estimated parameters. Similar performances are indicated by shape/color. This is the same as Figure 10 in the manuscript.}\label{fig:lin-principal-components2}
\end{figure}

In \autoref{sec:clust-music-perf} of the manuscript, \autoref{fig:pca}
showed the first two principal components along with some notion of
groups to gauge the similarities between performances. That figure is
reproduced in color as \autoref{fig:lin-principal-components2}.

For additional context, \autoref{tab:lin-pc-loadings} gives the loadings
for the first three principal components. We see that the first
component picks up information about the first two states, both through
\(\mu_{\textrm{tempo}}\) and \(\mu_{\textrm{acc}}\) as well as loading
onto the probabilities \(p_{11}\), \(p_{12}\), and \(p_{31}\). The
second component loads especially onto \(\mu_{\textrm{stress}}\) but
also \(p_{11}\) and \(p_{21}\). Finally, the third component loads
mainly onto the observation error with smaller contributions from
\(p_{31}\).

\begin{table}

\caption{\label{tab:lin-pc-loadings}The factor loadings for principal component analysis of the parameter estimates.}
\centering
\resizebox{\linewidth}{!}{
\begin{tabular}[t]{lrrrrrrrrrrrr}
\toprule
  & $\sigma_\epsilon^2$ & $\mu_{\textrm{tempo}}$ & $\mu_{\textrm{acc}}$ & $\mu_{\textrm{stress}}$ & $\sigma_{\textrm{tempo}}^2$ & $p_{11}$ & $p_{12}$ & $p_{31}$ & $p_{13}$ & $p_{21}$ & $p_{32}$ & $p_{22}$\\
\midrule
PC1 & 0.07 & 0.34 & -0.66 & -0.22 & -0.09 & 0.27 & -0.34 & 0.32 & 0.03 & -0.05 & -0.3 & -0.05\\
PC2 & 0.07 & 0.00 & -0.12 & -0.48 & -0.02 & -0.46 & 0.13 & -0.01 & 0.17 & 0.68 & 0.0 & -0.17\\
PC3 & 0.64 & -0.59 & 0.16 & 0.02 & -0.03 & 0.08 & -0.13 & 0.32 & 0.03 & 0.05 & -0.3 & -0.02\\
\bottomrule
\end{tabular}}
\end{table}

\hypertarget{confidence-intervals}{%
\subsection{Confidence intervals}\label{confidence-intervals}}

\begin{figure}

{\centering \includegraphics[width=5.333in,height=8in]{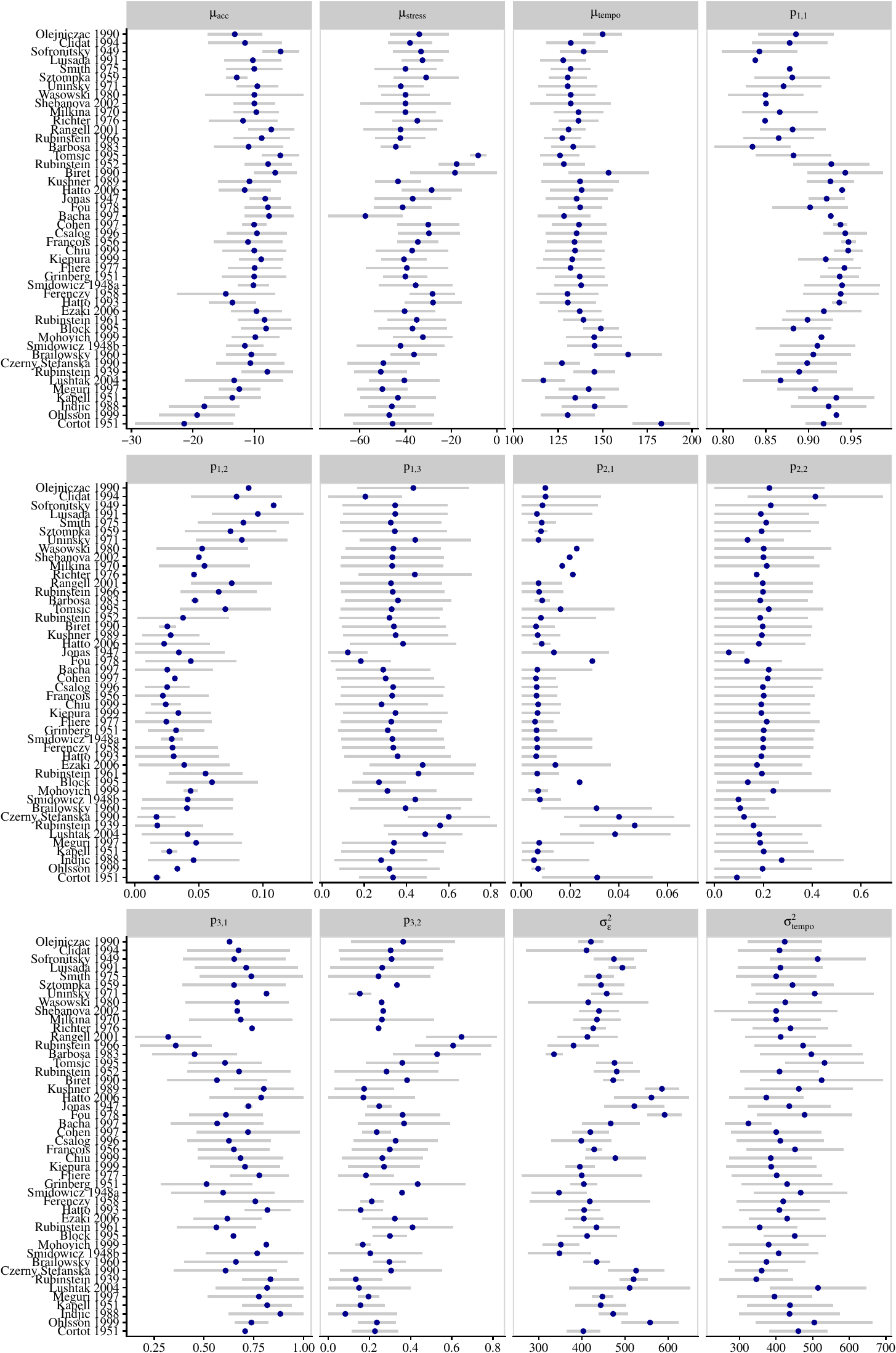} 

}

\caption{Confidence intervals for all parameters based on the observed Fisher information.}\label{fig:confidence-intervals}
\end{figure}

\begin{figure}

{\centering \includegraphics[height=2in]{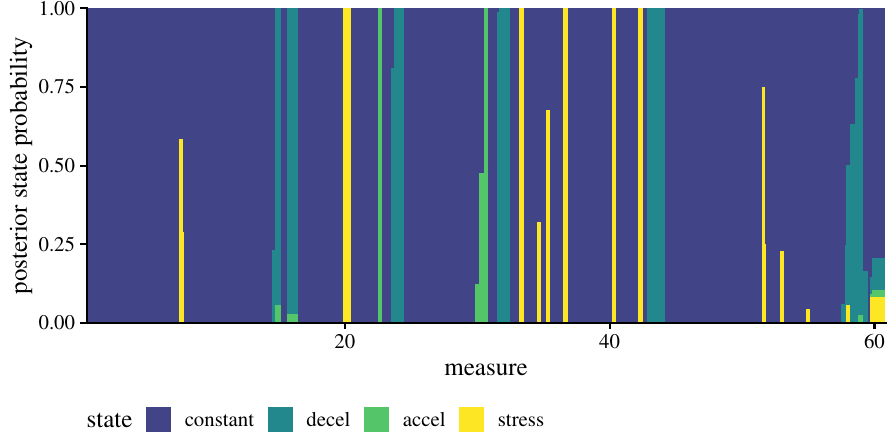} 

}

\caption{Distribution over potentitial states for Richter's 1976 recording.}\label{fig:posterior-richter-plot}
\end{figure}

\autoref{sec:analys-chop-mazurka} of the manuscript includes parameter
estimates for some of the recordings in our data set. In order to
quantify uncertainty and compare the estimates, this section graphically
displays all parameter estimates in \autoref{fig:confidence-intervals}.
The recordings are sorted in the same order as in \autoref{fig:dmats} in
the main document, so some of the conclusions about groupings are
readily apparent. The bars indicating measures of uncertainty are
derived form the observed Fisher information from the optimization
routine. However, it's not entirely clear what these mean. For one, they
ignore any uncertainty in the state sequence (see
\autoref{fig:posterior-richter-plot} some notion of the scale of this
uncertainty). They also depend on identifiability, the priors, and the
approximation to the posterior. Producing the MAP depends on the
approximation accuracy at the MAP, but producing the Hessian needs that
as well as the accuracy of about \(5p^2\) additional function
evaluations. And because we need the inverse, any inaccuracies could
explode. The length of the confidence interval for parameter \(j\) is
given by \(4\sqrt{(\widehat{I})^{-1}_{jj}}\) and so these would have
roughly 95\% coverage. For any parameters that are unidentified, the
width of the band is the maximum over all performances for the same
parameter. However, short of performing a fully-Bayesian analysis, we
would hesitate to attach much certainty to these metrics of uncertainty.

\hypertarget{distance-matrix-from-raw-data}{%
\subsection{Distance matrix from raw
data}\label{distance-matrix-from-raw-data}}

\begin{figure}[b]

{\centering \includegraphics[width=4in]{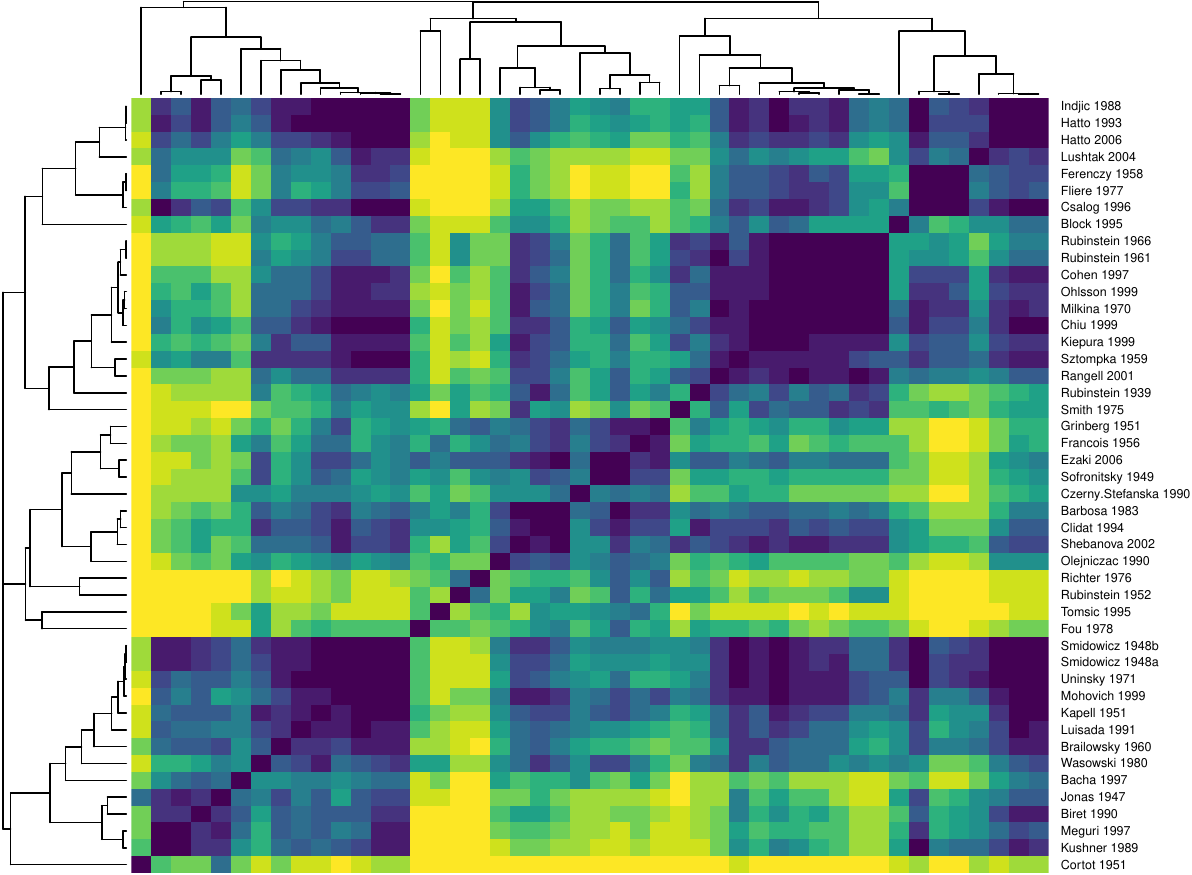} 

}

\caption{This figure presents a heatmap and hierarchical clustering based only on the note-by-note onset timings for each of the 46 recordings.}\label{fig:raw-data-clusters}
\end{figure}

In \autoref{sec:clust-music-perf} of the manuscript, we present results
for grouping performances using the low-dimensional vector of
performance-specific parameters learned for our model. An alternative
approach is to simply use the raw data, in this case, 231 individual
note-by-note instantaneous speeds measured in beats per minute. In
\autoref{fig:raw-data-clusters} we show the result of this analysis. A
comparison between this clustering and that given by our model is
discussed in some detail in the manuscript.

\hypertarget{plotting-performances}{%
\subsection{Plotting performances}\label{plotting-performances}}

\begin{figure}

{\centering \includegraphics{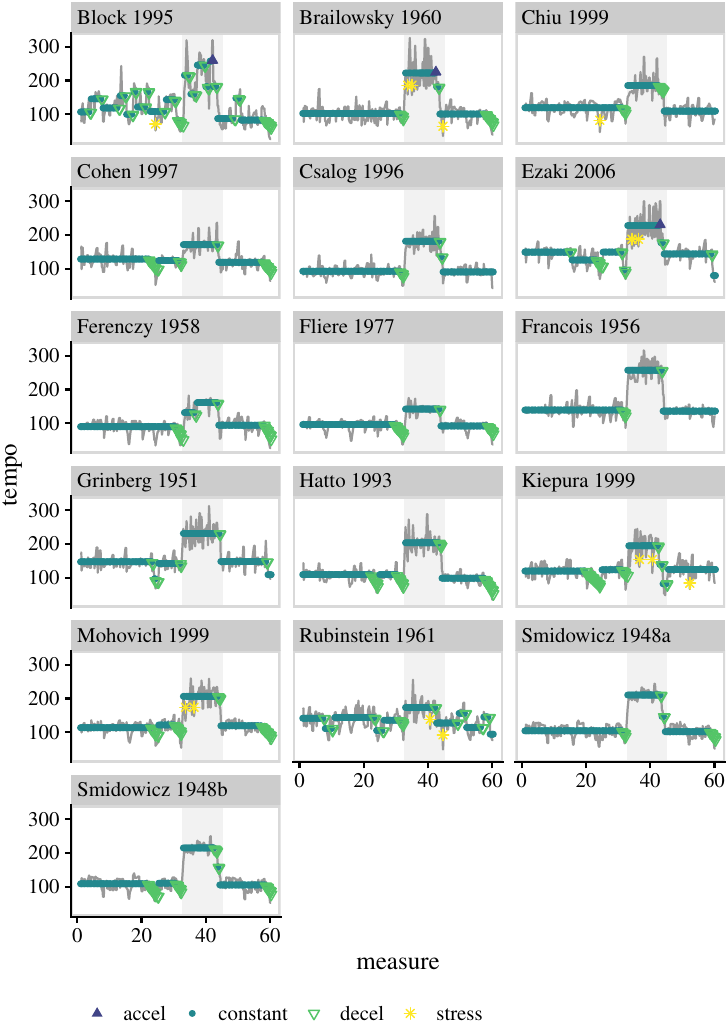} 

}

\caption{Performances in the first group}\label{fig:clust-1}
\end{figure}

\begin{figure}

{\centering \includegraphics{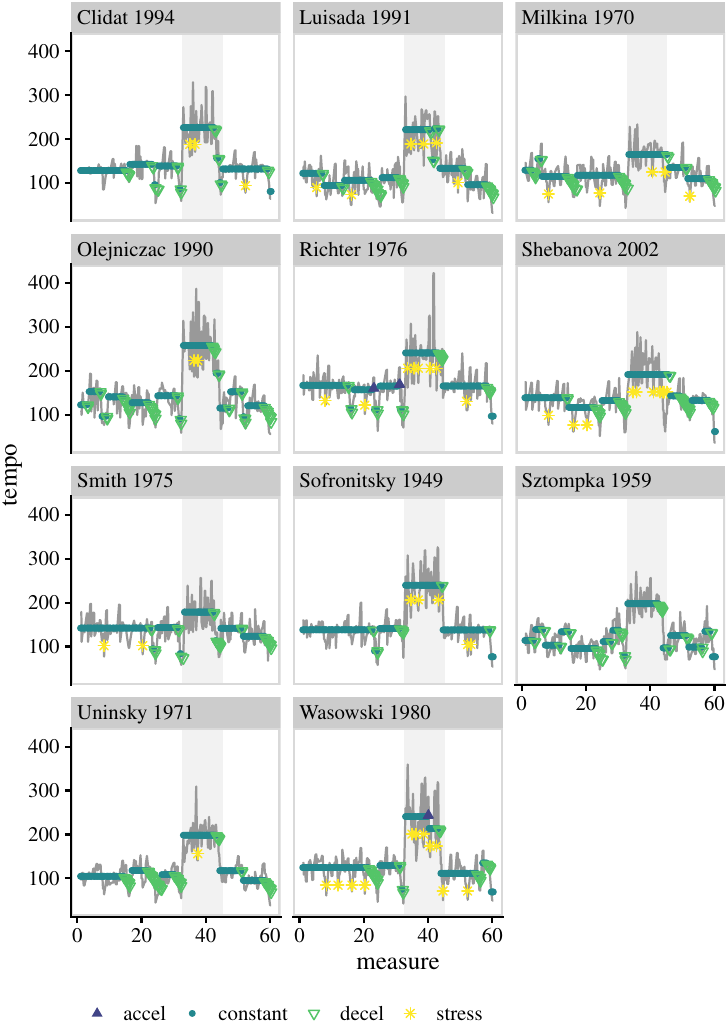} 

}

\caption{Performances in the fourth group.}\label{fig:clust-4}
\end{figure}

\begin{figure}

{\centering \includegraphics{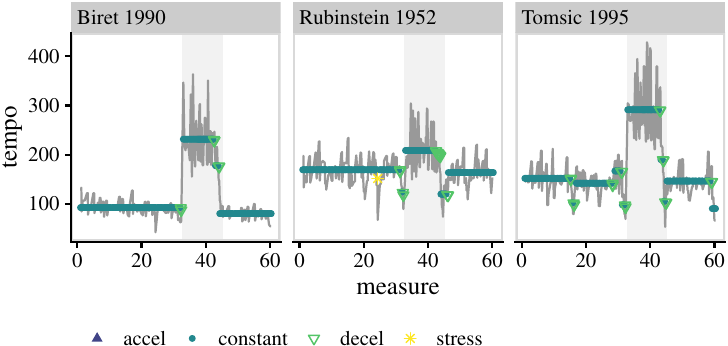} 

}

\caption{Performances in the sixth group.}\label{fig:clust-6}
\end{figure}

\begin{figure}

{\centering \includegraphics{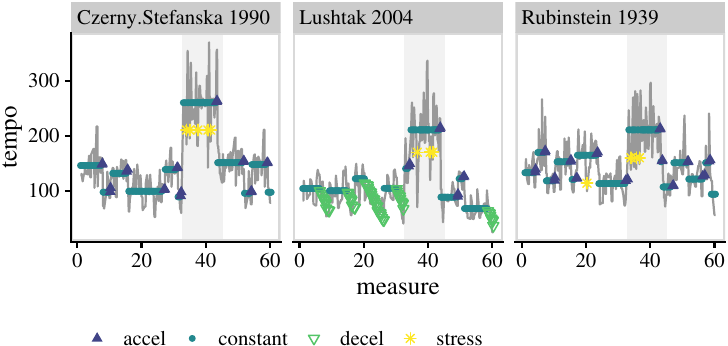} 

}

\caption{Performances in the second group.}\label{fig:clust-2}
\end{figure}

\begin{figure}

{\centering \includegraphics{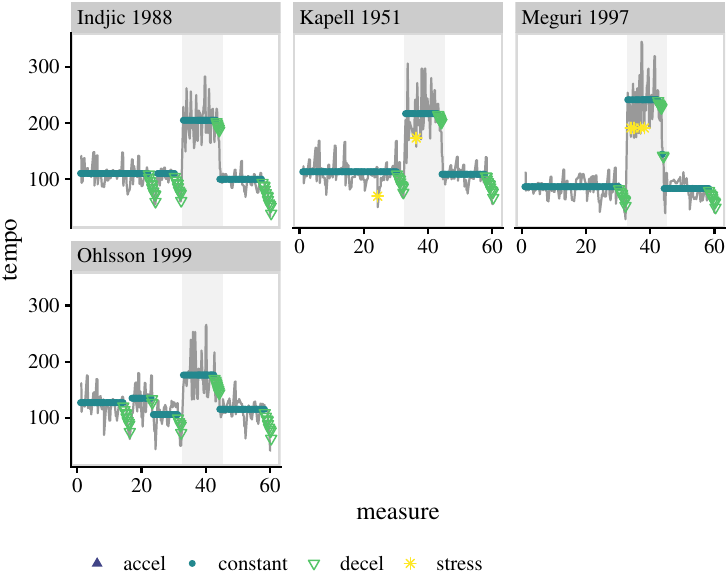} 

}

\caption{Performances in the third cluster.}\label{fig:clust-3}
\end{figure}

\begin{figure}

{\centering \includegraphics{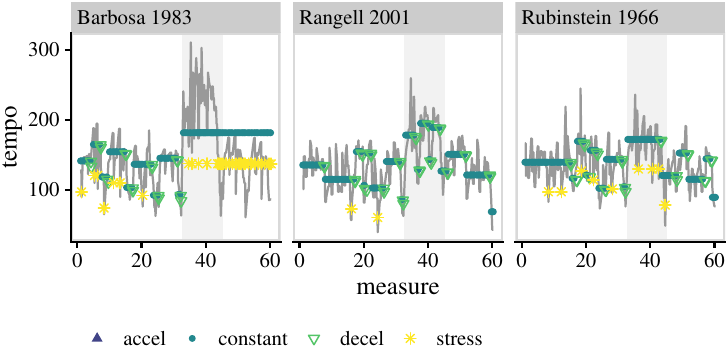} 

}

\caption{Performances in the fifth group.}\label{fig:clust-5}
\end{figure}

\begin{figure}

{\centering \includegraphics{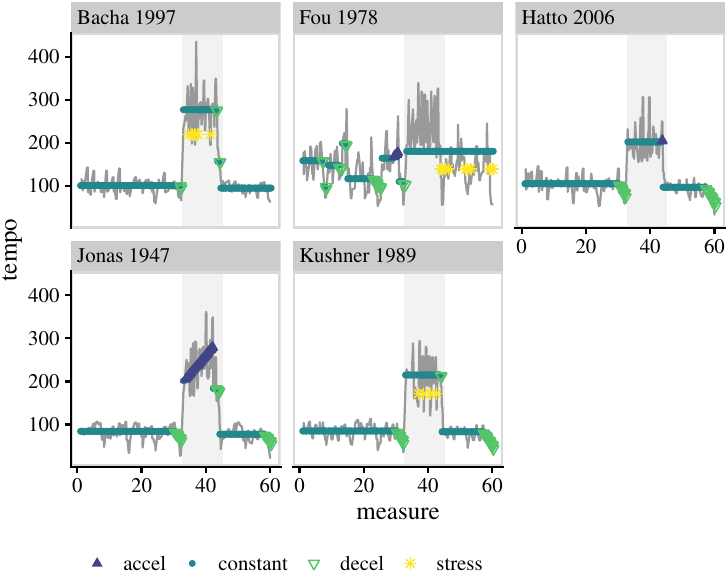} 

}

\caption{Performances in the seventh group.}\label{fig:clust-7}
\end{figure}

\autoref{sec:clust-music-perf} of the manuscript discusses 7 groups of
recordings. Figures \ref{fig:clust-1} to \ref{fig:clust-7} display the
note-by-note tempos along with the inferred interpretive decisions for
all performances based on this grouping.

The first group (\autoref{fig:clust-1} indicated as \(\circ\) in
\autoref{fig:pca}) corresponds to reasonably staid performances. This
group is the largest and corresponds to the block from Cohen to
Brailowsky in \autoref{fig:dmats}. In this group, the emphasis state is
rarely visited with the performer tending to stay in the constant tempo
state with periods of slowing down at the ends of phrases. Acceleration
is almost never used. Furthermore, these performances have relatively
slow average tempos, and not much difference between the A and B
sections. Joyce Hatto's recording in \autoref{fig:archetypal} is typical
of this group

Recordings in the fourth group (\autoref{fig:clust-4}, \(\oplus\) in
\autoref{fig:pca}) are those in the upper right of \autoref{fig:dmats},
from Olejniczac to Richter. These recordings tend to transition quickly
between states, especially constant tempo and slowing down accompanied
by frequent transitory emphases. The probability of remaining in state 1
is the lowest while the probability of entering state 2 from state 1 is
the highest. The acceleration state is rarely visited. Four of the most
similar performances are in this group along with Richter's 1976
recording.

The three performances in group six (\autoref{fig:clust-6}, \(\bullet\)
in \autoref{fig:pca}) are actually quite like others, but with small
exceptions. Biret's 1990 performance is very much like those in group 1,
but with a much larger contrast between tempos in the A and B sections.
The recording by Rubinstein in 1952 is similar, though with a faster A
section that has less contrast with the B section. Tomsic's 1995
performance is actually most similar to those in group three
(\(\mathrlap{+}\times\)), but played much faster and with a large
\(\sigma^2_\epsilon\).

The remaining performances are displayed in Figures
\ref{fig:clust-2}--\ref{fig:clust-7}, with the exception of Cortot's
performance in the manuscript.

\hypertarget{investigation-of-oversmoothing}{%
\subsection{Investigation of
oversmoothing}\label{investigation-of-oversmoothing}}

In \autoref{sec:this-model-reas} we argue that our model is not
eliminating too much tempo information that would increase listener
enjoyment. That is, we eliminate extraneous ``noise'' rather than
beneficial ``signal.'' \citet{GuRaphael2012} examine a related though
simpler model and undertake an empirical investigation by asking
experienced musicians to distinguish between recordings synthesized from
their tempo model and actual performances. They find that listeners were
not meaningfully able to distinguish between the two in the majority of
experiments, suggesting that their model retains most of the ``signal.''
Our model is more expressive than theirs, and therefore even less likely
to oversmooth. To examine this assertion, we compare the inferred tempo
decisions for their model and ours for 3 performances.

\begin{figure}

{\centering \includegraphics{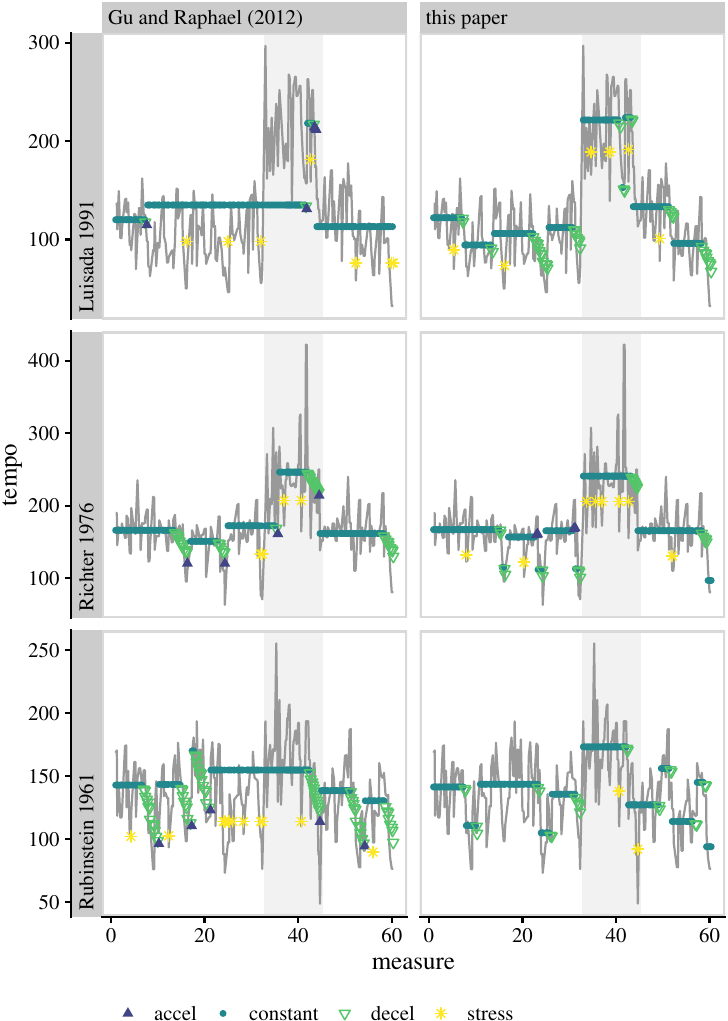} 

}

\caption{Inferred performance decisions for the tempo model in this paper compared to that examined in previous switching work.}\label{fig:yp-performances}
\end{figure}

\autoref{fig:yp-performances} shows the inferred tempo decisions for the
model in this paper in the right column and those produced by
\citet{GuRaphael2012} on the left. We use the same prior distributions
to estimate their model as described in the main document with minor
adjustments to eliminate two probabilities from the Markov switching
matrix. In the Luisada performance, our model is far more expressive,
allowing for many more changes of tempo. This is most easily seen in the
first A section where the \citet{GuRaphael2012} model gets stuck in the
constant tempo state, smoothing away all tempo information. It gets
similarly stuck in the Rubinstein recording, though there it appears to
potentially have some additional flexibility, in the frequent use of the
slowing-down state. Rather than using the linear ``slowing down''
behavior, our model begins a new constant tempo section in many of these
passages.

\begin{figure}

{\centering \includegraphics{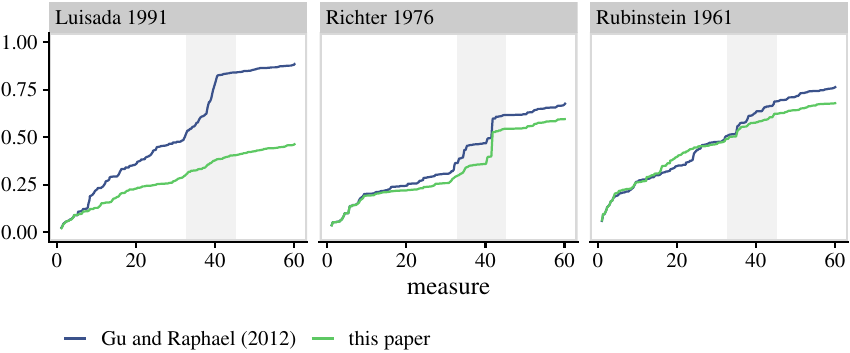} 

}

\caption{Cumulative RMSE over the recording relative to that of smoothing with the mean (something like $1-R^2$).}\label{fig:yp-residuals}
\end{figure}

To examine the amount of smoothing more specifically,
\autoref{fig:yp-residuals} compares the cumulative sums of the absolute
residuals over the recording relative to the cumulative sum of using the
mean. A value of 1 here indicates a completely over-smoothed recording
where we use the mean for a synthesized performance while a value of
zero corresponds to the performed tempos. Clearly, in each case, the
\citet{GuRaphael2012} model smooths more than the model presented in
this paper. Across all 46 performances, 35 are less smooth for our
model, and the relative absolute residuals are smaller on average 0.506
vs 0.614.

\hypertarget{distribution-over-states}{%
\subsection{Distribution over states}\label{distribution-over-states}}

To examine the stability of the \autoref{alg:dpf}, we examined all the
potential paths for Richter's 1976 recording. Here, we saved the most
likely 10,000 paths and their weights (rather than only the most likely
path). \autoref{fig:posterior-richter-plot} shows the marginal
(posterior) probability of being in a particular state for each note.
While the paper uses the most likely \emph{path}, this figure is
marginal in the sense that a particular note/state combination will have
high probability when many paths visited that note/state. But, the most
likely path may not have used that same note/state combination.
Nonetheless, there appears to be consensus for many of the notes. The
most obviously difficult notes are those near measures 10 and 50. In
both cases, the most likely path (\autoref{fig:richter} in the main
text) used the stress state, which exceeds 50\% posterior probability
here.

\hypertarget{multiplicative-tempo-changes}{%
\subsection{Multiplicative tempo
changes}\label{multiplicative-tempo-changes}}

\begin{figure}

{\centering \includegraphics[width=5in]{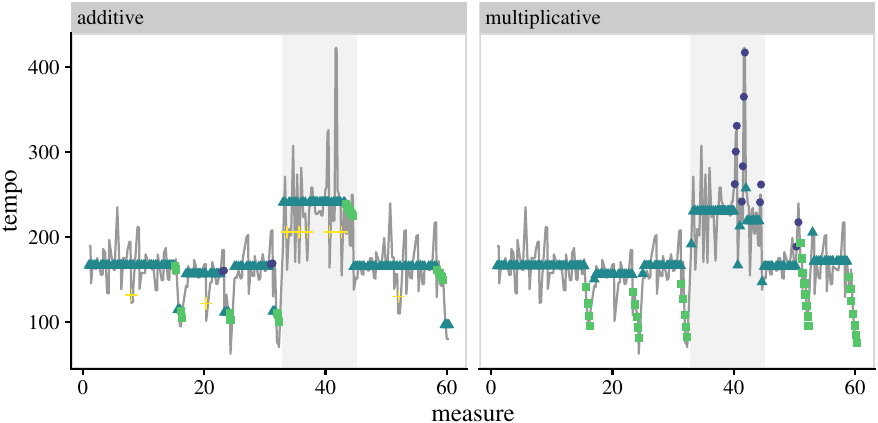} 

}

\caption{Additive and multiplicative models for Richter's 1976 performance. The multiplicative model fits quite well, but is less likely to visit the stress state.}\label{fig:multiplicative-model}
\end{figure}

While an additive state space model is relatively easy to understand,
some music theorists \citep[for example]{Mead2007} have argued that
musicians make multiplicative tempo adjustments. That is, it is the
ratio between the tempo of the current note and that of the previous
note rather than their difference that is important. Such a conception
is fundamental to musical notation (quarter notes, eighth notes, etc)
and frequently used to specify tempo changes within a piece of music,
such as with \halfnote~=~\quarternote~to indicate that the next section
should be played at half the previous tempo.

Rather than the linear switching model described in
\autoref{sec:materials-methods}, we examined the following
multiplicative version: \begin{equation}
  \begin{aligned}
    x_1 &\sim \textrm{lognormal}(x_0,\ P_0),\\
    \frac{x_{i+1}}{x_i} &= (1-\mu(s_i)) \eta_i, 
    & \eta_i &\sim \textrm{lognormal}(0,\ Q(s_i)),\\
    y_i&= c(s_i) +x_i + \epsilon_i, & \epsilon_i &\sim N(0,\ G(s_i)).
  \end{aligned}
\end{equation} Here, \(\mu(s_i)\) is 0 in the constant tempo or emphasis
states and controls the magnitude of acceleration or deceleration in the
other states. To complete the model,
\(G=\sigma^2_\epsilon + \sigma^2_{stress}I(s_i=4)\) while
\(Q = \sigma^2_{acc}\) in states 2 or 3 and 0 otherwise. To make this
model easier to compute, we transform by examining the log of the
transition equation and exponentiating the hidden continuous state in
the measurement equation. Likelihood evaluation is then performed with
the extended Kalman filter (EKF). The EKF is essentially the Kalman
filter applied to the first-order Taylor series expansion of any
non-linear components around our current predictions of them. For this
model, we have \begin{equation}
  \begin{aligned}
    \log(x_1) &\sim N(x_0,\ P_0),\\
    \log(x_{i+1}) &= \log(x_i) + \log(1-\mu(s_i)) + \log(\eta_i), 
    & \log(\eta_i) &\sim N(0,\ Q(s_i)),\\
    y_i&= c(s_i) +\exp(\varx_i) + \exp(\varx_i)x_i + \epsilon_i, & \epsilon_i &\sim N(0,\ G(s_i)),
  \end{aligned}
\end{equation} where \(\varx_i\) is the estimate of
\(E\left[x_i \given y_1,\ldots y_{i-1}\right]\).

We estimated this same model on the entire dataset and performed
principal component analysis on the resulting parameters (see also
\autoref{sec:clust-music-perf} of the manuscript).
\autoref{fig:multiplicative-model} shows the inferred performance
decisions for Richter's 1976 performance from both the linear and
multiplicative models. Both seem to fit the data quite well. There are
three slight differences between the inferences. First, in the B
section, the multiplicative model uses the acceleration state more than
the additive model does. Second, the multiplicative model avoids the
stress state, and this behavior is reflected in a higher probability of
remaining in the constant tempo state. Third, the perioods of slowing
down at the ends of phrases are better explained by the multiplicative
model.

\begin{figure}

{\centering \includegraphics[width=5in]{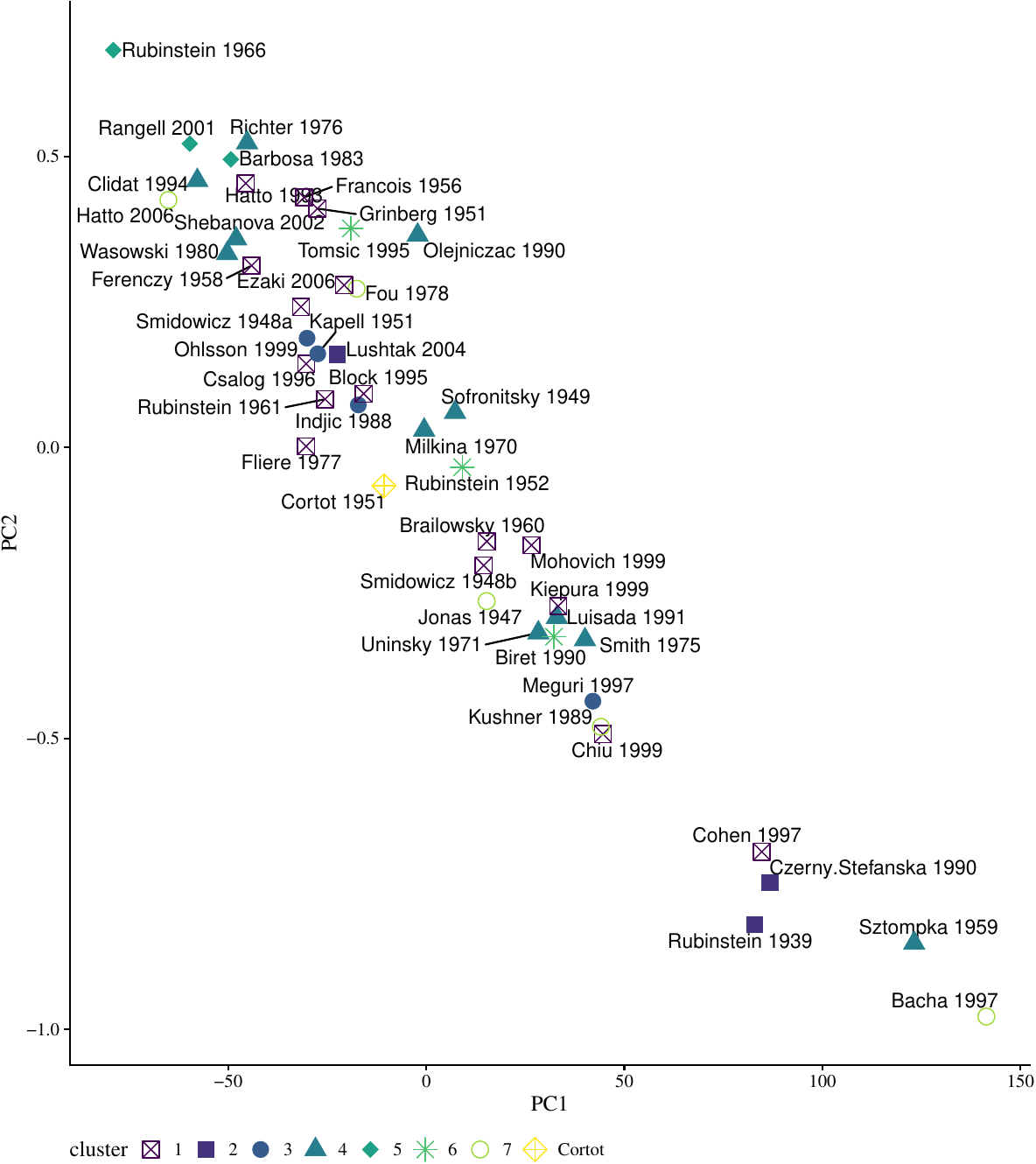} 

}

\caption{The first two principal components based on parameter estimates from the multiplicative model. The groupings (color and point type) are the same as those from the linear model.}\label{fig:mult-par-clusts}
\end{figure}

The percent of variance explained by the first two principal components
is 99\%. The first factor loads completely on \(\sigma^2_{\epsilon}\)
while the second loads on \(\mu_{\textrm{stress}}\). So, while this
model can explain individual performances quite well, it is much less
able to provide musically meaningful distinctions between performances.
While the interpretation for Richter's performance seems quite
reasonable under this model, other performances are much less
reasonable.

\hypertarget{alternative-prior-distributions}{%
\subsection{Alternative prior
distributions}\label{alternative-prior-distributions}}

Following the recommendations of an anonymous referee, we reestimated
the model under some alternative prior distributions. These
distributions are shown in \autoref{tab:morepriors}. Returning to
Richter's recording, we used inverse gamma and uniform distributions for
the variance parameters to allow heavier tails. We also looked at
uniform distributions on the transition probabilities and a prior which
requires the observation variance, \(\sigma^2_\epsilon\) to be smaller.

\begin{figure}

{\centering \includegraphics[width=5in]{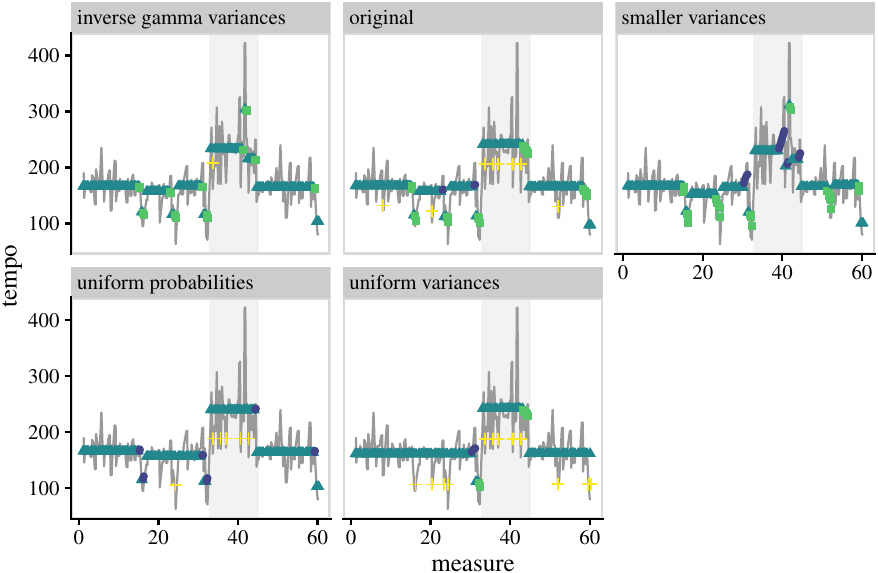} 

}

\caption{Inferred state sequence for Richter's 1976 recording under alternative prior specifications.}\label{fig:alternative-priors}
\end{figure}

Apart from the ``smaller observation variance'' setting, these different
specifications do not have a dramatic effect: the fit to the data
remains similar both quantitatively (as measured by RMSE and negative
loglikelihood, see \autoref{tab:prior-mses}) and qualitatively (as
determined by examining the inferred performance in
\autoref{fig:alternative-priors}).

The prior modes are important for some parameters to avoid
non-identifiability, and occasionally, as described in the manuscript,
to enforce more musically meaningful switching behaviors. On the other
hand, the prior tail shape is not particularly important here because
we're estimating posterior modes rather than performing a full Bayesian
analysis with accompanying credible intervals.

\begin{table}[t]
    \caption{Informative prior distributions for the music model}
  \label{tab:morepriors}
  \centering
  \begin{tabular}{@{}rcccc@{}}
    \toprule
    Parameter & \phantom{a} & Original & Inverse Gamma \\
    \midrule
    $\sigma^2_{\epsilon}$ & $\sim$ & Gamma$(40,\ 10)$ & IG$(42,16400)$\\
    $\mu_{\textrm{tempo}}$ & $\sim$ & Gamma$\left(\frac{\overline{Y}^2}{100},\ \frac{100}{\overline{Y}}\right)$ & IG$\left(\frac{\overline{Y}^2}{100}+2, \overline{Y}(\frac{\overline{Y}^2}{100}+1)\right)$\\
    $-\mu_{\textrm{acc}} $ & $\sim$ & Gamma$(15,\ 2/3)$ & IG(17, 160)\\
    $-\mu_{\textrm{stress}} $ & $\sim$ & Gamma$(20,\ 2)$ & IG(22, 840\\
    $\sigma^2_{\textrm{tempo}} $ & $\sim$ & Gamma$(40,\ 10)$ & IG$(42,16400)$\\
    $\sigma^2_{\textrm{acc}} $ & $=$ & 1 &  1\\
    $\sigma^2_{\textrm{stress}} $ & $=$ & 1 & 1\\
    $p_{1,\cdot}$ & $\sim$ & Dirichlet$(85,\ 5,\ 2,\ 8)$& Dirichlet$(85,\ 5,\ 2,\ 8)$ \\
    $p_{2,\cdot}$ & $\sim$ & Dirichlet$(4,\ 10,\ 1,\ 0)$& Dirichlet$(4,\ 10,\ 1,\ 0)$ \\
    $p_{3,\cdot}$ & $\sim$ & Dirichlet$(5,\ 3,\ 7,\ 0)$& Dirichlet$(5,\ 3,\ 7,\ 0)$ \\
    \midrule
    Parameter & \phantom{a} & Smaller $\sigma^2_\epsilon$ &
    Uniform Variances & Uniform Probabilities\\
    \midrule
    $\sigma^2_{\epsilon}$ & $\sim$ & Gamma$(20,\ 10)$ & 1 & Gamma$(20,\ 10)$ \\
    $\mu_{\textrm{tempo}}$ & $\sim$ & Gamma$\left(\frac{\overline{Y}^2}{100},\ \frac{100}{\overline{Y}}\right)$ & Gamma$\left(\frac{\overline{Y}^2}{100},\ \frac{100}{\overline{Y}}\right)$ & Gamma$\left(\frac{\overline{Y}^2}{100},\ \frac{100}{\overline{Y}}\right)$ \\
    $-\mu_{\textrm{acc}} $ & $\sim$ & Gamma$(15,\ 2/3)$ & Gamma$(15,\ 2/3)$ & Gamma$(15,\ 2/3)$\\
    $-\mu_{\textrm{stress}} $ & $\sim$ & Gamma$(20,\ 1)$ & Gamma$(20,\ 2)$ & Gamma$(20,\ 2)$ \\
    $\sigma^2_{\textrm{tempo}} $ & $\sim$ & Gamma$(40,\ 10)$ & 1 & Gamma$(40,\ 10)$\\
    $\sigma^2_{\textrm{acc}} $ & $=$ & 1 &  1 & 1\\
    $\sigma^2_{\textrm{stress}} $ & $=$ & 1 & 1 & 1\\
    $p_{1,\cdot}$ & $\sim$ & Dirichlet$(85,\ 5,\ 2,\ 8)$& Dirichlet$(85,\ 5,\ 2,\ 8)$& 1 \\
    $p_{2,\cdot}$ & $\sim$ & Dirichlet$(4,\ 10,\ 1,\ 0)$& Dirichlet$(4,\ 10,\ 1,\ 0)$& 1 \\
    $p_{3,\cdot}$ & $\sim$ & Dirichlet$(5,\ 3,\ 7,\ 0)$& Dirichlet$(5,\ 3,\ 7,\ 0)$& 1 \\
    \bottomrule
  \end{tabular}
\end{table}

\begin{table}

\caption{\label{tab:prior-mses}For each of the different prior distributions, we report the MSE between the estimated performer intentions from the model and the true performance as well as the negative log likelihood.}
\centering
\begin{tabular}[t]{lrrr}
\toprule
prior & rmse & negative log likelihood & $\sigma_{\epsilon}$\\
\midrule
inverse gamma variances & 28.57 & -5.20 & 20.66\\
original & 29.31 & -5.10 & 19.24\\
smaller variances & 26.77 & -5.10 & 20.01\\
uniform probabilities & 30.54 & -5.06 & 21.27\\
uniform variances & 31.05 & -5.15 & 23.56\\
\bottomrule
\end{tabular}
\end{table}

\begin{table}

\caption{\label{tab:posterior-ests}For each of the different prior distributions, we report the estimated parameter values.}
\centering
\resizebox{\linewidth}{!}{
\begin{tabular}[t]{lrrrrr}
\toprule
  & original & smaller variances & inverse gamma variances & uniform variances & uniform probabilities\\
\midrule
$\sigma^2_{\epsilon}$ & 426.70 & 370.37 & 400.36 & 452.49 & 555.29\\
$\mu_{\textrm{tempo}}$ & 136.33 & 167.65 & 166.55 & 133.34 & 135.69\\
$\mu_{\textrm{acc}}$ & -11.84 & -23.55 & -6.44 & -10.02 & -7.74\\
$\mu_{\textrm{stress}}$ & -34.82 & -16.76 & -24.75 & -54.16 & -50.89\\
$\sigma^2_{\textrm{tempo}}$ & 439.38 & 451.58 & 400.57 & 406.50 & 320.17\\
$p_{11}$ & 0.85 & 0.94 & 0.91 & 0.90 & 0.93\\
$p_{12}$ & 0.05 & 0.03 & 0.05 & 0.02 & 0.01\\
$p_{22}$ & 0.74 & 0.46 & 0.52 & 0.68 & 0.85\\
$p_{31}$ & 0.44 & 0.22 & 0.30 & 0.29 & 0.74\\
$p_{13}$ & 0.02 & 0.01 & 0.01 & 0.01 & 0.03\\
$p_{21}$ & 0.25 & 0.47 & 0.45 & 0.26 & 0.08\\
$p_{32}$ & 0.17 & 0.04 & 0.20 & 0.13 & 0.15\\
\bottomrule
\end{tabular}}
\end{table}

\end{document}